\documentclass{article}

\usepackage{PRIMEarxiv}

\usepackage{ragged2e} 
\usepackage{booktabs,makecell, multirow, tabularx}

\usepackage[utf8]{inputenc} 
\usepackage[T1]{fontenc}    
\usepackage[colorlinks=true, allcolors=blue]{hyperref}
\usepackage{hyperref}       
\usepackage{url}            
\usepackage{booktabs}       
\usepackage{amsfonts}       
\usepackage{nicefrac}       
\usepackage{microtype}      
\usepackage{lipsum}
\usepackage{fancyhdr}       
\usepackage{graphicx}       
\graphicspath{{media/}}     
\usepackage{amsmath}
\usepackage{subcaption}
\usepackage{float}
\title{Data Driven Topology Optimization of Channel Flow Problems
}

\author{
  Ce Guan, Jianyu Zhang, Zhen Li, Yongbo Deng \\
  State Key Laboratory of Applied Optics \\
  Changchun Institute of Optics, Fine Mechanics and Physics (CIOMP) Chinese Academy of Sciences \\
  Changchun 130033, China\\
  \texttt{dengyb@ciomp.ac.cn} \\
}

\begin{document}

\maketitle

\begin{abstract}
Typical topology optimization methods require complex iterative calculations, which cannot meet the requirements of fast computing applications. The neural network is studied to reduce the time of computing the optimization result, however, the data-driven method for fluid topology optimization is less of discussion. This paper intends to introduce a neural network architecture that avoids time-consuming iterative processes and has a strong generalization ability for topology optimization for Stokes flow. Different neural network methods including Convolution Neural Networks (CNN), conditional Generative Adversarial Networks (cGAN), and Denoising Diffusion Implicit Models (DDIM) which have been already successfully used in solid structure optimization problems are mutated and examined for fluid topology optimization cases. The presented neural network method is tested on the channel flow topology optimization problems for Stokes flow. The results have shown that our presented method has high pixel accuracy, and gains a 663 times decrease on average in execution time compared with the conventional method.
\end{abstract}

\date{Version \versionno, \today}

\maketitle 


\section{Introduction}

Topology optimization is a mathematical method to place material within a prescribed design domain in order to obtain a certain system performance by setting an objective function subject to some constraints\cite{Intro-1-browne2013topology}.  By using topology optimization, the unnecessary features or material in the production is reduced, and the cost of time required for design is also greatly reduced. Topology optimization has been a focus of research for many years since proposed by Bendsøe et al. in 1988 \cite{9-1-bendsoe1988generating}.  It was first used in mechanical design problems and has been expanded to other physical fields, such as fluids, acoustics, electromagnetics, optics, and their combinations \cite{9-2-sigmund2013topology}.

For topology optimization of fluid flow problems, the initial investigation for topology optimization by the density method for fluidic problems was studied for Stokes flows by Borrvall and Petersson \cite{5-borrvall2003topology}. A design parameterization, which depends on density controlling permeability, was introduced by Borrvall and Petersson, to identify the non-fluid places and fluid places \cite{5-borrvall2003topology}.
Followed by other topology optimization problems on Stokes flow\cite{5-1-guillaume2004topological, 5-2-aage2008topology}, the parameterization idea was extended to solve topology optimization problems for Navier-Stokes flow in 2005 by Gersborg-Hansen et al. \cite{9-4-gersborg2005topology} Follow-up studies have improved the topological optimization for Navier-Stokes flows in both methodology\cite{Intro-3-guest2006topology, Intro-4-wiker2007topology}, implementation\cite{Intro-2-olesen2006high, Intro-7-othmer2008continuous}, and parametrization\cite{Intro-5-bruns2007topology}. It was later spread to other more intractable flows including non-Newtonian fluiData Set \cite{9-6-pingen2010optimal, Intro-6-ejlebjerg2012topology}, unsteady flows\cite{9-5-deng2011topology, Intro-11-deng2013topology}, and turbulence flows\cite{Intro-8-yoon2016topology, Intro-9-dilgen2018density, Intro-10-yoon2020topology}.

Many methodologies for solving topology optimization have been investigated, including the density-based method (i.e., the solid isotropic material with penalization (SIMP) method), the evolutionary structural optimization method, and the level-set method. More recently, for fluid problems, the methodology for particular applications has been studied. For example, the parametric level-set method has been applied to fluid topology optimization problems using non-uniform mesh \cite{10-2-li2022topology}.

These conventional methods are intended to increase the accuracy and stability of the computation, however, they do not avoid iterations. The high computational cost is one of the main challenges in traditional topology optimization procedures \cite{0-0-shin2022topology}, which obtains the optimal result by computing sensitivity for a large number of iterations, in which each iteration performs the time-consuming finite element analysis (FEA) to calculate the sensitivity.

In recent decades, machine learning, and especially neural network methods have developed rapidly and have been successfully applied to many fielData Set \cite{0-0-shin2022topology}. For topology optimization, in order to reduce the time and space complexity of topology optimization, many machine learning-based topology optimization methods have been investigated. Machine learning can improve the existing topology optimization techniques in many ways, including acceleration of iteration, non-iterative optimization, meta-modeling, dimensionality reduction of design space, improvement of the optimizer, generative design (design exploration), and post-processing \cite{0-0-shin2022topology}. 

Research on two-dimensional structure optimization using machine learning methods has been widely studied, including traditional shapes (beam, bridge, truss)\cite{3-TopoCNN-wang2022deep, 1-TopoGAN-nie2021topologygan},  and more typical applications (Wheel \cite{6-21-oh2018design}, Motor \cite{6-20-sasaki2019topology}). Other than topology optimization for structures, the data-driven method has also been used in several other physical fields. Deep learning methods have been used for topology optimization to design the geometry of photonic devices \cite{6-13-liu2018training, 6-11-mao2023multi, 6-12-kojima2021inverse}. Li. et al. have ultimately employed the deep learning models to design phononic crystals with anticipated band gaps \cite{6-14-li2020designing}. The optimal heat conduction system is also studied to design through deep learning method \cite{6-15-deng2022self, 6-16-keshavarzzadeh2021image, 6-17-zhang2021tonr}. In terms of fluid problems, Yaji et al. used machine learning algorithms to construct high-fidelity results through the procedure of low-fidelity modeling \cite{fluid_based1}; Mahammod et al. improved the modeling accuracy of Reynolds stress in the turbulence model through the machine learning method, thereby enhancing the performance of RANS-based turbulence topology optimization \cite{fluid_based2}. 

In this study, we present the data-driven topology optimization framework for the channel flow problems, which resolves the problem that fluid topology optimization is too slow in execution time. The neural network architectures are used to build the mapping between the physical fields computed from the unoptimized design domain to the optimized fluid structures. To test our presented framework, we tried three neural network architectures, including convolution neural network (CNN), conditional generative adversarial network (cGAN), and denoising diffusion implicit models (DDIM). Unlike the traditional topology optimization method for Stokes flow which requires a large number of iterations, the presented data-driven method topology optimization method for the channel flow problems can calculate the optimized result directly from the unoptimized physical fields. The main aim of our presented method is to reduce the computational cost of traditional topology optimization for Stokes problems. Indeed, the execution time of neural network-based topology optimization is negligible compared to that of traditional topology optimization for Stokes flows.

The rest of this paper is organized as follows: In Section 2, we give a brief background on the density-based topology optimization method and three types of typical generative neural networks we used in this study. The general frameworks of neural network-based fluid topology optimization are described in section 3. We introduce 2 data sets that are generated for testing the neural network framework, and we show the performance of three neural network methods on 2 different data sets respectively in Section 4. Following that, we conduct comparative studies between the results from the density-based method and those generated through neural networks. These comparisons include pixel-wise error comparison (i.e., MSE and MAE),  subjective comparisons of optimal design and execution time comparison.

Our main contributions are as follows:  $(1)$ two new data sets for channel Stokes flow are generated for solving topology optimization problems based on the neural network; $(2)$ a neural network framework that computes the optimal design for fluid topology optimization directly from the physical fields over the initial design domain, which significantly reduces computation time for fluid topology optimization, is presented. 
\section{Methodology}
In this part, we will first briefly introduce the typical method for topology optimization for Stokes flow problems. Then, we will review the machine learning methods for topology optimization. Following that,  we will give an overview of the architecture of different generative neural networks.
\subsection{Typical Topology Optimization for Stokes Fluid Problems}
In this study, the presented method is to solve the classical minimization dissipation problem for Stokes flow, while, in principle, other objective functions or other types of flow problems should also be feasible using this framework. Considering the design domain $\Omega$, the goal of the topology optimization for Stokes flow is to find an optimal subset of $\Omega$, $\Omega_{Optimal} \subset \Omega$, that minimize the performance for an objective function $f(y, \alpha)$ as in \eqref{eq::OptGen}, under constraints of given conditions in the design domain $\Omega$, as in \eqref{eq::OptGen}, where the design variable $\alpha$ determines the subset of the design domain $\Omega$, and $y$ is the physical fields (velocity $\mathbf{u}$ and pressure $p$) corresponding to the given $\alpha$.
\begin{equation}\label{eq::OptGen}
\begin{aligned}
& \min_\alpha:  \quad \quad  f(y, \alpha) \\
& \text {subject to}\left\{\begin{array}{l}
\text{Stokes equations for $y$ and $\alpha$ in $\Omega$} \\
\text{Boundary Conditions for $y$ on $\partial \Omega$} \\
\text{Conditional constrains for design variable $\alpha$ in $\Omega$}
\end{array}\right. \\
\end{aligned}
\end{equation}
The dynamic properties of velocity $\boldsymbol{u}$ and pressure $p$ in the computational domain satisfy the Stokes equations. The velocity Dirichlet conditions are given at the boundary of the design domain. Considering the objective is to minimize the power dissipation in the design domain,  subject to Stokes equation under the viscosity setting to 1, the topology optimization problem for the Stokes flow can be formulated as follows:
\begin{equation}\label{eq::FluidOpt}
\begin{aligned}
 \min : & \quad \frac{1}{2} \int_{\Omega} \alpha \boldsymbol{u} \cdot \boldsymbol{u}+\nabla \boldsymbol{u}: \nabla \boldsymbol{u}-\int_{\Omega} f u \\
 \text { s.t. } & -\nabla \cdot(\nabla \boldsymbol{u}-p I)+\alpha \boldsymbol{u}=0 \quad \text { in } \Omega \\
& \quad \nabla \cdot \boldsymbol{u}=0 \quad \text { in } \Omega \\
& \boldsymbol{u} = \boldsymbol{u}_D \quad \text { in } 
 \partial \Omega \\
& \quad 0 \leq \rho \leq 1 \text { in } \Omega \\
& \quad \int_{\Omega} \rho \leq V \cdot|\Omega|
\end{aligned}
\end{equation}
where $\boldsymbol{u}$ is the fluidic velocity, $p$ is the fluidic pressure, $\alpha$ is density-based design variable, denoted as a function of  $\rho$, that is $\alpha(\rho)$,  ($\rho$ is the control parameter, $\rho = 1$ means fluids present, $\rho = 0$ means solid present), $f$ is the prescribed source term 
(in this problem we take it to be $0$),
$\partial \Omega$ is the boundary of the design domain $\Omega$, 
and $\boldsymbol{u}_D$
is the Dirichlet boundary condition for velocity $\boldsymbol{u}$. $V$ is the volume fraction, the upper boundary of the control, and $|\Omega|$ is the volume of the whole design domain.
The design variable $\alpha(\rho)$ is modeled as \eqref{eq::DesignVariable}:

\begin{equation}\label{eq::DesignVariable}
\alpha(\rho)=\bar{\alpha}+(\underline{\alpha}-\bar{\alpha}) \rho \frac{1+q}{\rho+q}
\end{equation}
where $\bar{\alpha}$, $\underline{\alpha}$ 
and $q$ are constants, $\bar{\alpha}$ and $\underline{\alpha}$ 
are the minimal and maximal values of $\alpha$, and $q$ is the penalization deviation taken from the values 0 or 1. 

\subsection{Machine Learning Methods for Topology Optimization}
The machine learning(ML) technique was first used in topology optimization in a seminal paper proposed by  Ulu et al. \cite{6-4-ulu2016data}, which uses principal component analysis (PCA) to represent the optimal topologies and the neural network was built to map from loading information to PCA weights. After that, many other machine learning ideas are suggested. Lei et al. have employed supported vector regression (SVR) as well as K-nearest-neighbors (KNN)  to build a mapping from design parameters and loading information to optimal topology structure \cite{6-3-lei2019machine}. Long-Short Term Memory (LSTM) architecture is used for topology optimization by Qiu et al. to learn the structural evolution process from training data \cite{6-18-qiu2021deep}. The kriging model based on the support vector machine (SVM) for predicting the sensitivity is introduced by Xia et al. to reduce the computational cost for topology optimization \cite{6-19-xia2017new}.

The optimization procedure for finding optimal results for topology optimization might be considered as a generalization task. Thus,  the most commonly used strategy is by using encoder-decoder architectures, which is a very successful neural network architecture for generation tasks. The encoder extracts features from the input information and the decoder integrates the features obtained from the encoder to generate the expected result. Sosnovik et al. studied to use convolution 
encoder-decoder architecture to accelerate the topology optimization procedure \cite{6-1-sosnovik2019neural}, which speeds up topology optimization execution efficiency by reducing the number of iterations. In particular, they used the result from fewer iterations for the solid isotropic material penalization (SIMP) method as the input of the neural network, and  subsequent iterations are replaced by maps trained by the neural network, while the final result is the output of the neural network. A 3D encoder-decoder Convolution Neural Network(CNN) architecture is designed to generate the 3D topology optimization results in \cite{6-2-banga20183d}, which used density and loading information as input and 3D topology optimization generated by SIMP as the output. Recently, Wang et al. have proposed a U-net encoder-decoder deep convolution neural network, which has  higher accuracy compared with similar results from other CNNs, since they used physical fields as the input channels for the neural network, which contains more physical information \cite{3-TopoCNN-wang2022deep}. 

Generative adversarial network(GAN), as one of the most powerful neural network architectures, has also been widely studied for topology optimization tasks. Yu et al. have used the cGAN to construct high-resolution optimized structures from low-resolution ones \cite{6-5-yu2019deep}. Super-resolution generative adversarial network (SRGAN) has been suggested by Li. et al. to optimize heat transfer structure problems \cite{6-6-li2019non}. Oh et al. proposed a deep generative adversarial framework that can produce a large number of aesthetically pleasing and engineering performance-optimized design choices \cite{6-7-oh2019deep}. 
Rawat et al. came up with a convolution neural network (CNN) and a Wasserstein GAN into a coupled model for 3D topology optimization \cite{6-8-rawat2019application}. Nie et al. have proposed a cGAN-structured neural network framework, TopologyGAN, for solving structured topology optimization, which instantly predicts the optimized structure, aided by physics information computed from the original and unoptimized design domain through the finite element method \cite{1-TopoGAN-nie2021topologygan}. 
More recently, Maze et al. have proposed TopoDiff, which uses diffusion models on topology optimization problems and corrected the neglected manufacturability of the cGAN model \cite{6-9-maze2022diffusion}.\\
\subsection{UNet and CNN}
Convolutional neural networks (CNN) are a class of feed-forward neural networks that learn features through convolutional filters. In particular, features of input images can be learned automatically through convolution operation by extracting features through the convolution operator. The UNet is a convolution network with a U-shape encoder-decoder architecture proposed by Ronneberger et al. \cite{3-2-Unet-ronneberger2015u}. UNet was first proposed to solve the problem of medical image segmentation and further extended to other semantic segmentation tasks. This is because the UNet model is successfully trained by only fewer samples, and outperforms other models. The structure of UNet consists of two parts, the encoding part, and the decoding part. The purpose of the encoder is to capture contextual information for the input image through repeated convolution desample blocks. Meanwhile, the purpose of the decoder part is to use the transpose convolution upsampling blocks repeatedly to gradually restore feature maps to the desired resolution.

\subsection{cGAN\label{Sec::cGANIntro}}
\begin{figure*} 
    \centering
    \includegraphics[width=14cm]{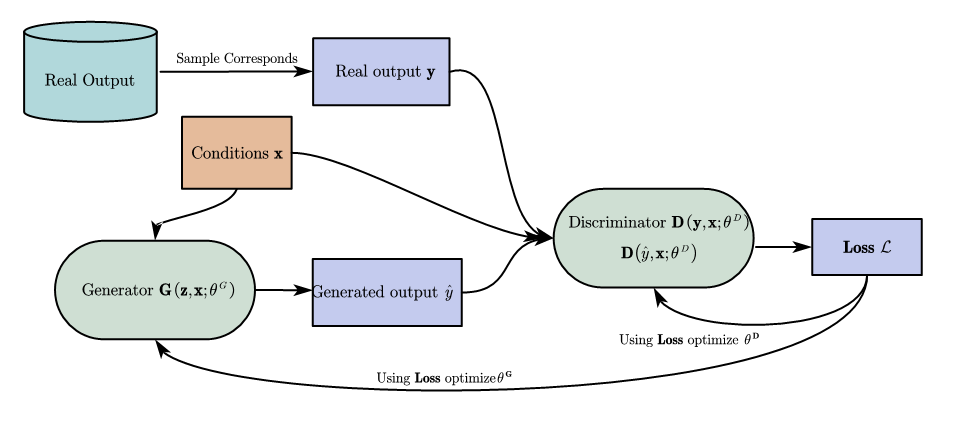}
    \caption{The architecture of cGAN}
    \label{fig::cGAN}
\end{figure*}

The generative adversarial nets (GAN) \cite{1-1-GAN-goodfellow2014generative}, is a novel way to train generative models. The cGAN is the conditional version of GAN \cite{1-2-cGAN-mirza2014conditional}, which allows setting conditions on the generator and discriminator. The architecture of cGAN is shown in Figure \ref{fig::cGAN}. The cGAN is a generative model that trains a mapping from random noise, which is denoted as $\boldsymbol{z}$ in Figure \ref{fig::cGAN}, and conditioning information, which is denoted as $\boldsymbol{x}$ in Figure \ref{fig::cGAN}, to the output result $\boldsymbol{\hat{y}}$.  The task of the generator $G(\boldsymbol{z}, \boldsymbol{x}; \boldsymbol{\theta}^G)$ is to receive conditions $\boldsymbol{x}$ and then use a neural network to create an output $\boldsymbol{\hat{y}} = G(\boldsymbol{z}, \boldsymbol{x}; \boldsymbol{\theta}^G)$ as close to the real output $\boldsymbol{y}$.
The discriminator $D(\boldsymbol{y}, \boldsymbol{x}; \boldsymbol{\theta}^D)$ is for learning to classify the generated output $\boldsymbol{\hat{y}}$ (fake output) and real output $\boldsymbol{y}$, while the generator $G$ is learned to fool the discriminator $D$. The discriminator in cGAN is a discriminating convolution neural network, and the task of the discriminator is to determine whether the input information of the discriminator is originated from $\boldsymbol{\hat{y}}$, which is generated by generator $G$, or from the sample $\boldsymbol{y}$. Since $D$ is used to determine whether the output is real or fake, the input of discriminator $D$ is set to be $(\boldsymbol{x}, \boldsymbol{\hat{y}} \textit{ or } \boldsymbol{y})$. The output of the discriminator $D$ is a real number, which determines whether the input of $D$ is fake (generated by G), or real (ground truth from the data set).
The cGAN iterates parameter $\boldsymbol{\theta}^G$ and $\boldsymbol{\theta}^D$ through loss function $\mathcal{L}$ to find the optimal model. $\mathcal{L}$  contains the loss function for the generator and the loss function for the discriminator. The loss function $\mathcal{L}$, should lower the loss functions for generator G, to let the output of the Generator be as close as the real ones in order for the generator to fool the discriminator as much as possible. However,  $\mathcal{L}$ is also intended to increase the loss function of the discriminator through training, so that the discriminator cannot distinguish between the result generated by the generator and the real result.  Therefore, the loss function for cGAN can be denoted as:
\begin{multline}\label{Original_cGAN_Loss}
\mathcal{L}_{c G A N}(G, D)=\mathbb{E}_{(x, y) \sim p_{\text {data }}(x, y)}[\log D(x, y)]+ \\ 
\mathbb{E}_{x \sim p_{\text {data }}(x), z \sim p_z(z)}[\log (1-D(x, G(x, z)))]
\end{multline}
After that, a similar framework with the pix2pix architecture \cite{1-3-isola2017image} was proposed, which uses U-Net as the generator \cite{3-2-Unet-ronneberger2015u}, and it does not need to add a noise $z$ in the generator. Thus, for pix2pix architecture, the generator $G$ is the mapping from information $x$ to the conditioning information $G(x)$, where $G(x)$ is the generated result from the generator.
\subsection{Diffusion Model}
The diffusion model is one of the most popular frameworks for generative neural networks. The diffusion model was first proposed in 2015 \cite{Dif-0-cD-sohl2015deep}. Ho et al. used denoising diffusion probabilistic models(DDPM) to present high-quality image synthesis results\cite{Dif-1-cD-ho2020denoising}. Song et al. have presented the denoising diffusion implicit model(DDIM) to accelerate the sampling of the DDPM model, which produces high-quality samples much faster\cite{Dif-3-song2020denoising}. Nichol further improved the diffusion and obtained much better state-of-art conditional image synthesis quality on the task\cite{Dif-2-cD-dhariwal2021diffusion} by adding guidance to the model. The principle of the diffusion model is to generate a mapping from the white noise distribution to the generated result, which is a reverse noising process, in which the task is to train a neural network to gradually remove noise from a signal. The diffusion model gradually adds noise to the sample that is the Gaussian noise is added gradually one step after another, and at the final step, the white noise distribution is obtained. In particular, consider the sample result as $x_0$, $x_t$ is the result of adding Gaussian noise from $x_{t-1}$, hence, each step produces noisier samples, up to $T$ steps where $x_T$ is the white noise. The forward  gradual Gaussian noising process is a Markov process and can be represented as:
\begin{equation}
q\left(\mathrm{x}_t \mid \mathrm{x}_{t-1}\right):=\mathcal{N}\left(\mathrm{x}_t ; \sqrt{1-\beta_t} \mathrm{x}_{t-1}, \beta_t \mathrm{I}\right)
\end{equation}
where $\beta_t$ is the variance schedule. According to the property, the sample result at arbitrary timestep can be obtained from the starting point, that is:
\begin{equation}
q\left(x_t \mid x_0\right)=\mathcal{N}\left(x_t ; \sqrt{\bar{\alpha}_t} x_0,\left(1-\bar{\alpha}_t\right) \mathbf{I}\right)
\end{equation}
where $\alpha_t=1-\beta_t$ and 
$\bar{\alpha}_t =\prod_{s=0}^t \alpha_s$.\\
The diffusion model is intended to train a neural network to reverse such a process, in particular, a denoising process, removing Gaussian noise from $x_t$ to $x_{t-1}$. The parametric Gaussian process can be represented as:
\begin{equation} \label{Diff_back_Train}
p_\theta\left(x_{t-1} \mid x_t\right):=\mathcal{N}\left(x_{t-1} ; \mu_\theta\left(x_t, t\right), \Sigma_\theta\left(x_t, t\right)\right)
\end{equation}
where the $\mu_\theta\left(x_t, t\right)$ and $\Sigma_\theta\left(x_t, t\right)$ are the mean and variance for the denoising process, and they are needed to be trained through the neural network. Note that, the forward process is fixed, and the $q\left(x_{t-1} \mid x_t, x_0\right)$ can be determined:
\begin{equation} \label{Diff_foward_known}
q\left(\mathbf{x}_{t-1} \mid \mathbf{x}_t, \mathbf{x}_0\right)=\mathcal{N}\left(\mathbf{x}_{t-1} ; \tilde{\boldsymbol{\mu}}_t\left(\mathbf{x}_t, \mathbf{x}_0\right), \tilde{\beta}_t \mathbf{I}\right),
\end{equation}
where $\quad \tilde{\mu}_t\left(\mathrm{x}_t, \mathrm{x}_0\right):=\frac{\sqrt{\bar{\alpha}_{t-1}} \beta_t}{1-\bar{\alpha}_t} \mathrm{x}_0+\frac{\sqrt{\alpha_t}\left(1-\bar{\alpha}_{t-1}\right)}{1-\bar{\alpha}_t} \mathrm{x}_t \quad$ and $\quad \tilde{\beta}_t:=\frac{1-\bar{\alpha}_{t-1}}{1-\bar{\alpha}_t} \beta_t$. 
The idea of the diffusion model is to use the neural network to train the mean and variance of the denoising steps \eqref{Diff_back_Train}. 
\section{Presented Method \label{Sec::Methods}}
\begin{figure}
    \centering
    \includegraphics[width=11.3cm]{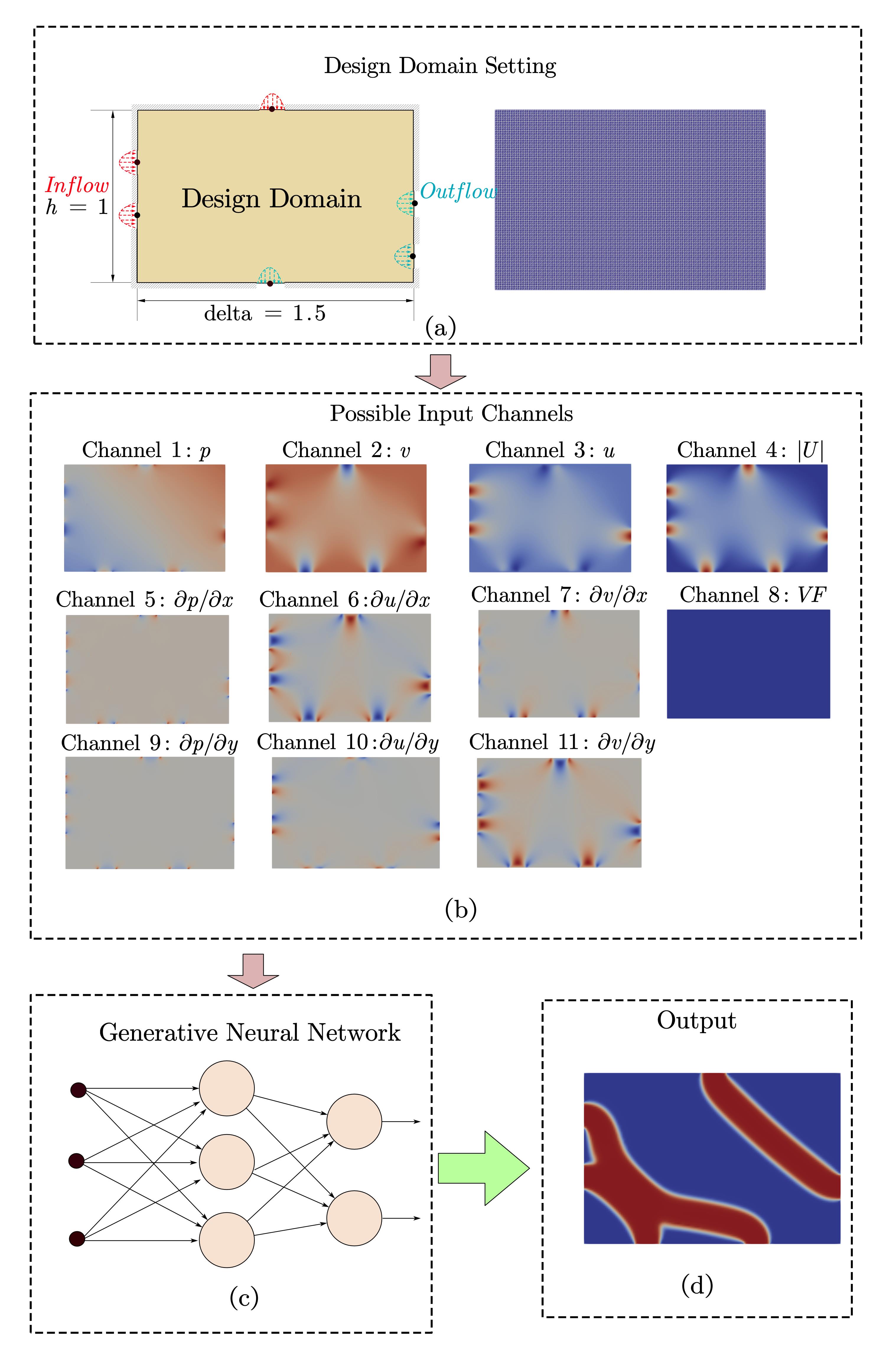}
    \caption{The conceptual flow of the presented mothod}
    \label{fig::NN}
\end{figure}

The conceptual flow of the presented method is shown in Figure \ref{fig::NN}.  The first step is to randomly set up the initial design domain, and the random boundary conditions on the boundary of the domain. The design domain is shown in Figure \ref{fig::NN} $(a)$, which is discretized by $100 \times 100$ rectangle elements. Then, the fluidic problem is solved in the original non-optimized design domain,  where the physical fields in the design domain are computed through FEA. In particular, the basic flow features including velocity fields and pressure fields are obtained  by simulating Stokes equation in the design domain. Following that, the various physical fields are calculated, such as spatial gradients for basic flow features and volume fraction. All features are collected as the possible input channels of the presented neural network, as shown in Figure \ref{fig::NN} $(b)$.
Finally, the presented neural network as shown in Figure \ref{fig::NN} $(c)$ maps the physical field, that we generated for the non-optimized domain in Figure \ref{fig::NN} $(b)$, to the optimal result for the fluid structure as shown in Figure \ref{fig::NN} $(d)$.\\
In contrast to traditional fluid topology optimization methods, which use multiple iterations of the FEA to reach the final fluid structure, the presented method requires only one FEA (when generating the input information to the neural network), which significantly reduces the time complexity of generating the optimal fluid structure.
The presented data-driven method of fluid topology optimization is based on neural-network methods for structure topology optimization frameworks which use physical fields as the input of the network, including TopologyGAN\cite{1-TopoGAN-nie2021topologygan}, deep CNN for topology optimization \cite{3-TopoCNN-wang2022deep}, and TopoDiff\cite{6-9-maze2022diffusion} framework. The samples are generated through the finite element method (FEM) using FEniCS and density-based topology optimization using Dolfin Adjoint\cite{2-mitusch2019dolfin}. Analogous to the physical fields input used in structural optimization in TopologyGAN\cite{1-TopoGAN-nie2021topologygan}, deep CNN for topology optimization \cite{3-TopoCNN-wang2022deep}, and TopoDiff\cite{6-9-maze2022diffusion} (as a fluid problem, also received inspiration from Xu et al. \cite{8-1-xu2021machine}) we present to use the velocity and pressure fields of the flow field and their spatial gradients as the input channels for the neural network. The basic flow features (including velocity fields and pressure fields) and the possible channels(including spatial gradients of basic flow features and volume fraction) are shown in Table \ref{table::designvariables}\\
\begin{table*}
    \centering
    \caption{Flow-field Features}
\begin{tabular}{c|c}
\hline Basic flow features & $(u, v, p)$ \\
\hline $\begin{array}{c}\text { Features including spatial } \\
\text { gradients }\end{array}$ & $\left(Volume Fraction, |U|,\frac{\partial u}{\partial x}, \frac{\partial u}{\partial y}, \frac{\partial v}{\partial x}, \frac{\partial v}{\partial y}, \frac{\partial p}{\partial x}, \frac{\partial p}{\partial y}\right)$ \\
\hline
\end{tabular}
    \label{table::designvariables}
\end{table*}

\subsection{CNN Framework for Fluid Topology Optimization\label{Sec::CNN}}
The first neural network framework we intend to use for addressing fluid topology optimization is CNN.
Lots of the architectures for generative CNN had been tested by Zhang et al. \cite{3-TopoCNN-wang2022deep}, including GoogLeNet\cite{7-1-szegedy2015going}, ResNet \cite{7-2-he2016deep} and UNet\cite{7-3-ronneberger2015u} for structure optimization, and they find out that the UNet framework can significantly improve the pixel accuracy compared to other CNNs. Therefore, drawing inspiration from their research, we use UNet framework as the CNN framework for fluid topology optimization.\\

\begin{figure}
    \centering
    \includegraphics[width=16cm]{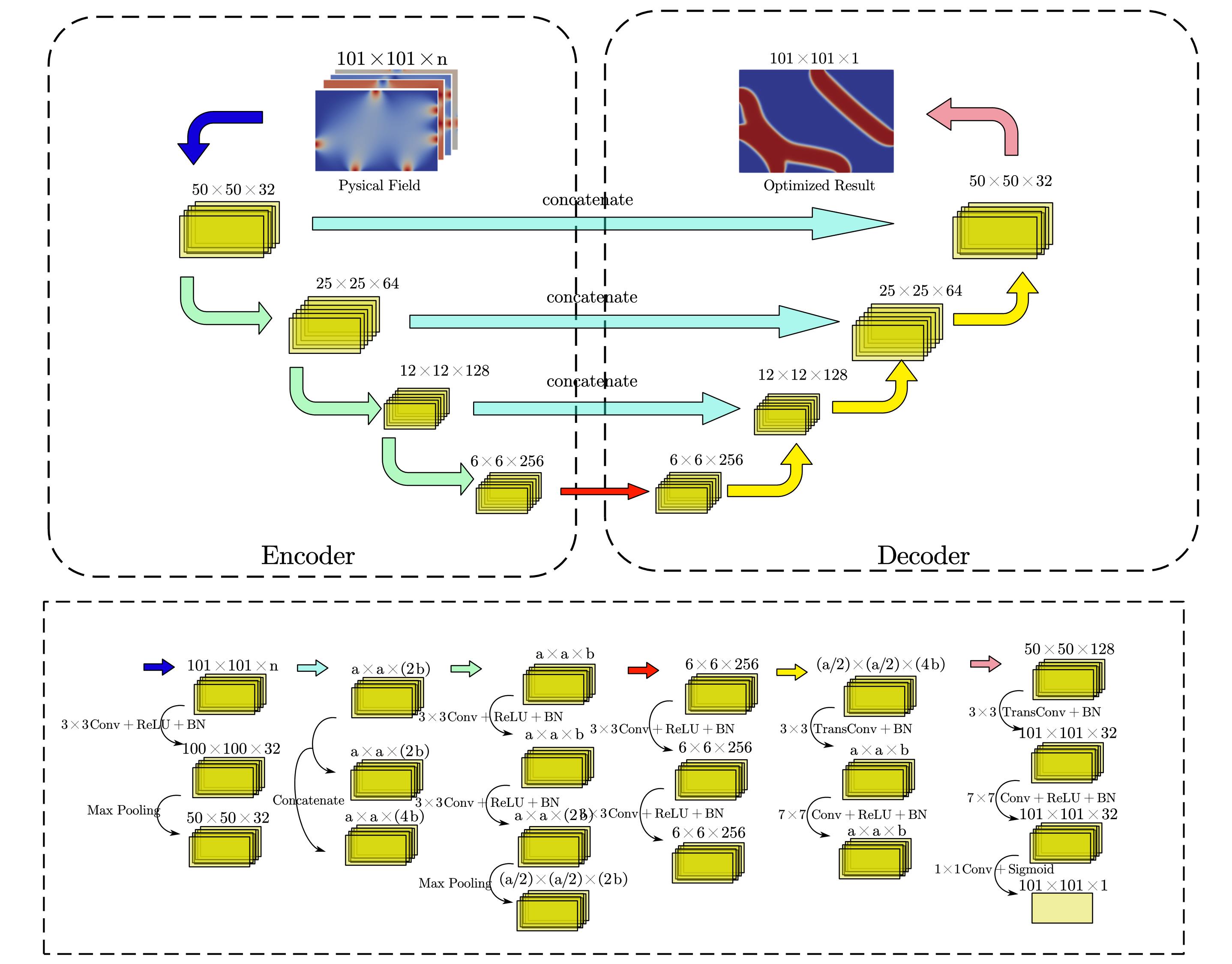}
    \caption{The architecture of CNN model for fluid-based topology optimization}
    \label{fig::CNN}
\end{figure}
The architecture of the CNN for solving the fluid topology optimization is shown in Figure \ref{fig::CNN}. For the encoder part, the objective is to down-sample the given array and return a dimension-reduced array. For the decoder part, the objective is to up-sample the given array and return a dimension ascended array \cite{3-TopoCNN-wang2022deep}. In the encoder,  from the beginning, the input tensor (physical fields channels from Figure \ref{fig::GAN_Experiment} $(b)$ ) with shape $101 \times 101 \times n$ is sent to the convolution layer, following by a $3 \times 3$ convolution and a $2 \times 2$ max pooling layer, which is shown as the purple arrow in Figure \ref{fig::CNN}. Then comes the three down-sampling steps, which are shown as green arrows in Figure \ref{fig::CNN}. For each down-sampling step, it consists of applying two $3 \times 3$ convolutions repeatedly. A ReLU and batch normalization are followed by each convolution. After that, the $2 \times 2$ max pooling layer is used for reducing dimensions. The down-sampling step doubles the channel of the tensor and reduces the dimension to half.  
There are 3 up-sampling steps for the decoder part, which are shown as yellow arrows in Figure \ref{fig::CNN}. The upsampling is detailed as follows: the input tensor is first concatenated with the corresponding feature map from the contracting path, following by a $3 \times 3$ transpose convolutions(with ReLu activation and batch normalization), which double the dimensions and reduce the channels. After that, a $7 \times 7$ convolution layer is used. The final layer, which is shown as pink arrow, is an up-sampling block following by a $1 \times 1$ convolution to map the channels to the final result with dimension $101 \times 101 \times 1$.

\subsection{cGAN Framework for Fluid Topology Optimization}
\begin{figure*}
    \centering
    \includegraphics[width=18cm]{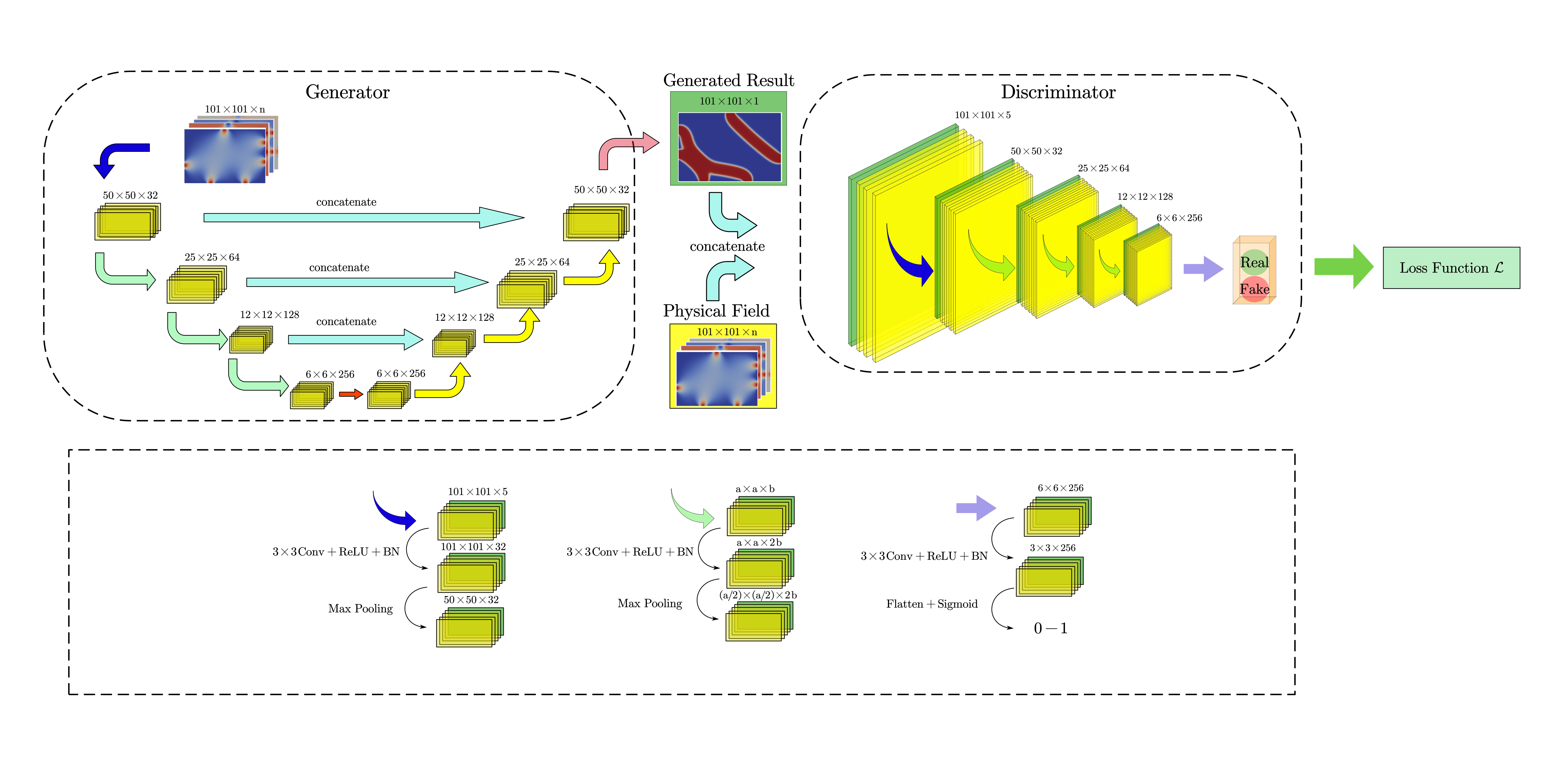}
    \caption{The architecture of cGAN model for fluid-based topology optimization}
    \label{fig::GAN_Experiment}
\end{figure*}

The second neural network framework we try to use in this study is the cGAN. The detailed architecture for the cGAN framework for fluid topology optimization is shown in Figure \ref{fig::GAN_Experiment}. The architecture of cGAN has already been introduced in the preceding Section \ref{Sec::cGANIntro}. The generator of cGAN for fluid topology optimization is the UNet structure, which is same as we used in the CNN framework for solving fluid topology optimization \ref{Sec::CNN}.
The discriminator, as shown in Figure \ref{fig::GAN_Experiment}, 
 for each green arrow in the discriminator, uses modules with the convolution-BatchNorm-ReLu with 1/2 dropout for each block \cite{9-3-ioffe2015batch}. The final layer, which is the purple arrow in the discriminator in Figure \ref{fig::GAN_Experiment}, is a fully connected layer with data flattened. Finally, the sigmoid function is used to classify between fake and real.

\subsection{Diffusion Model framework for Fluid Topology Optimization}
\begin{figure}
    \centering
    \includegraphics[width=14.5cm]{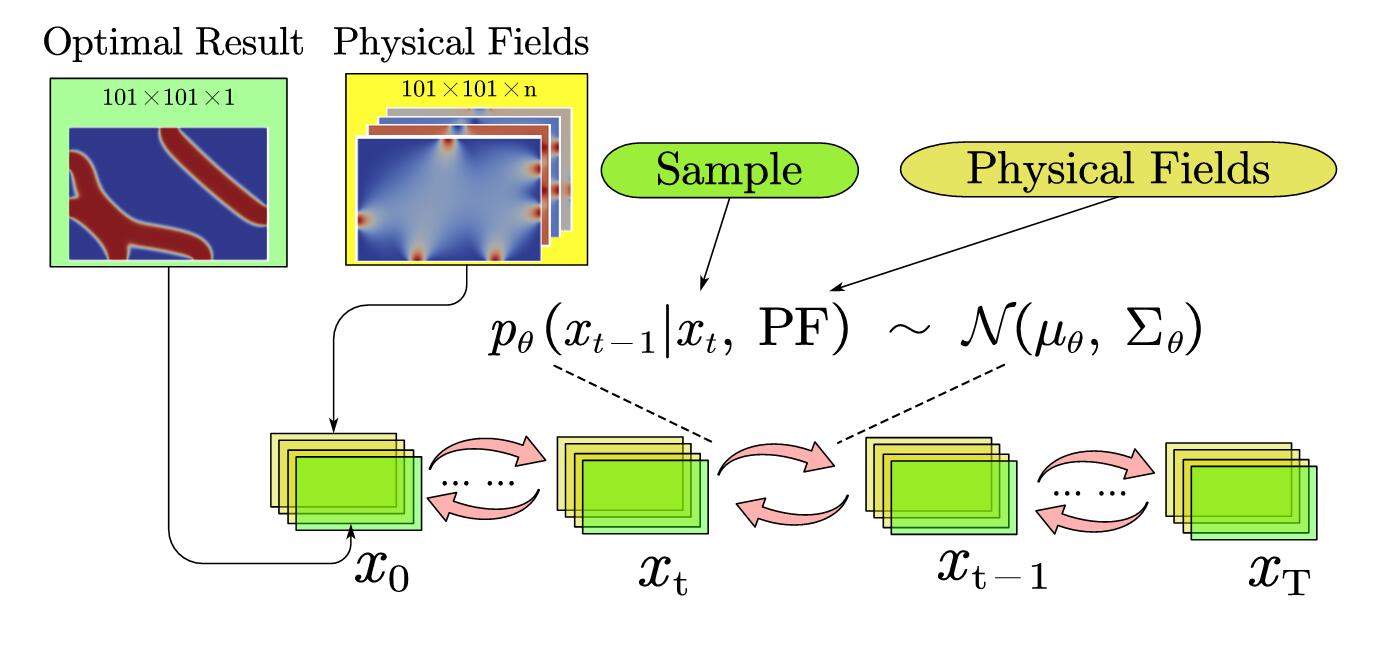}
    \caption{The architecture of DDIM model for fluid-based topology optimization}
    \label{fig::DDIM}
\end{figure}

\begin{figure*}\label{AddNoise}
    \centering
    \includegraphics[width=14cm]{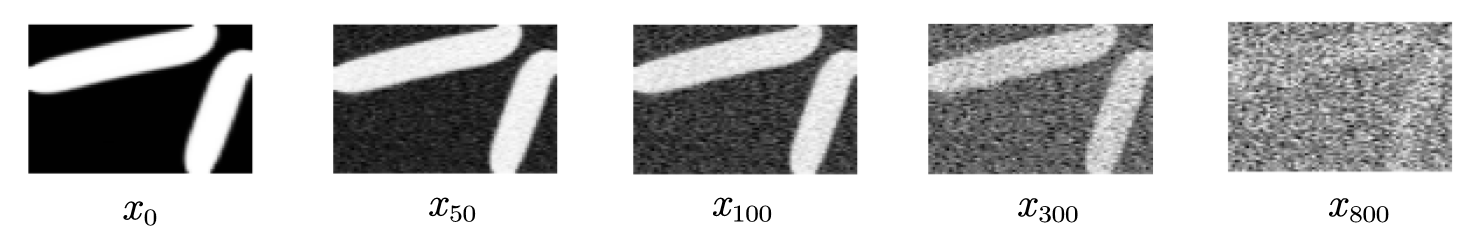}
    \caption{An example of the forward procedure of adding noise to a sample}
    \label{fig::AddNoiceSample}
\end{figure*}

The third neural network framework we used for solving data-driven fluid topology optimization is the diffusion model framework DDIM. The diffusion model framework for topology optimization for fluid problems is similar has already been used for solid topology optimization in \cite{6-9-maze2022diffusion}. In \cite{6-9-maze2022diffusion}, Francois and Faez presented a conditional diffusion-model-based architecture called TopoDiff, which uses physical fields as conditions and adds the conditions to the input image of the denoiser at each time step.
The general framework is shown in Figure \ref{fig::DDIM}, where $x_0$ is the optimal result as in the green box, and the conditional information $PF$ is fluid physical fields and volume fractions as in the yellow box. Note that, we train the diffusion model by training the variance at each denoising step using UNet, and extra channels as additional information are added to the input of UNet. In particular, the input of UNet is $x_t$, $t$, and $PF$, and the output of the UNet is the predicted noise added at time $t$. An example of the forward procedure of adding noise to a sample is shown in Figure \ref{fig::AddNoiceSample}.


\subsection{Loss Functions of CNN for Fluid Topology Optimization}
The loss function for the CNN-based framework consists of two parts: the pixel-wise error $\mathcal{L}(PW)$, and the volume fraction error $\mathcal{L}(VF)$, which is represented as follows:
\begin{equation}\label{Loss_CNN}
\mathcal{L}_{\text{CNN}} = \mathcal{L}(PW) + \lambda \mathcal{L}(VF)
\end{equation}
where,  $\lambda$ is the scale parameter used to balance these two errors. $\mathcal{L}(PW)$ is the loss of pixel-wise error between the optimal design from the density-based approach, and the optimized structure predicted through neural networks. Similar to the concept of UNet used in image segmentation, our objective is to solve a binary image segmentation problem for images (separating the design domain into two parts: fluid and solid). Thus, similar to UNet for image segmentation, which uses cross-entropy as the loss function, we also use cross-entropy as our $\mathcal{L}(PW)$, that is:
\begin{equation}\label{Loss_org_CNN}
\mathcal{L}(PW)  =\frac{1}{M} \sum_{i = 1}^ {M} \mathbf{y}_i \log \frac{\mathbf{y}_i}{\hat{\mathbf{y}}_i}
\end{equation}
where $M$ is the number of samples, $\mathbf{y}_i$ is the $i$th sample, and $\hat{\mathbf{y}_i}$ is the generated result through UNet. $\mathcal{L}(VF)$ is the penalty term for volume fraction, which is the error of volume fraction between the predicted result $\hat{\mathbf{y}_i}$ and density-based result $\mathbf{y}_i$. The $\mathcal{L}{(VF)}$ is identified as the $L_1$ loss, and it is shown as: 
\begin{equation}\label{VF_representation}
    \mathcal{L}{(VF)} = \frac{1}{M}\sum_{i = 1}^{M} \left|VF_{y_i} - VF_{\hat{y}_i}\right| =\frac{1}{M}\sum_{i = 1}^{M} \frac{1}{N} \left|\sum_{e = 1}^N y_{ie} - \sum_{e = 1}^N\hat{y}_{ie} \right|
\end{equation}
where $N$ is the number of elements in each sample, and  $y_{ie}$ represents the n design variable of the element $e$ in sample $i$.  $y_{ie}$ ranges from 0 to 1 in the topology optimization problem, where $y_{ie} = 0$ corresponds to an artificial solid domain and $y_{ie} = 1$ to a fluidic domain, respectively. The summation of $y_{ie}$ over a sample $i$ can represent the constituent of the fluidic domain divided by the volume of all design domains, which is the volume fraction.

\subsection{Loss Functions of cGAN for Fluid Topology Optimization}
Referring to the work done by Nie et al. on structure optimization using the cGAN framework \cite{1-TopoGAN-nie2021topologygan} in which the loss function adds the L2 loss of the generator and the absolute error of the generator's volume fraction as corrections to the typical cGAN loss, as shown in \eqref{Original_cGAN_Loss}, we set the loss function of our cGAN for fluid topology optimization as follows:
\begin{equation}
G^*=\arg \max _D \min _G \mathcal{L}_{cGAN}(G, D) +\lambda_1 \mathcal{L}_{L 2}(G)+\lambda_2 \mathcal{L}(VF)
\end{equation}
where $\mathcal{L}_{cGAN}$ is the loss function of cGAN neural network,   $G$ is the generator, $D$ is the discriminator, $\mathcal{L}_{L 2}$ and $\mathcal{L}(VF)$ are the $L2$ loss of the generator and $L1$ error for volume fraction, $\lambda_1$ and  $\lambda_2$ are the coefficients which are used to balance each part of the loss. The loss function  $\mathcal{L}_{cGAN}(G, D)$  is given by:
\begin{multline}
\mathcal{L}_{cGAN}(G, D)=-\frac{1}{M} \sum_{i=1}^{M} \log \mathrm{D}\left(D(\mathbf{x}_i, \mathbf{y}_i)\right) \\
-\frac{1}{M} \sum_{i=1}^{M} \log \left(1-\mathrm{D}\left(1- D(\mathbf{x}_i, G(\mathbf{x}_i))\right)\right)
\end{multline}
the $\mathcal{L}_{L2}(G)$ is computed by:
\begin{equation}
\mathcal{L}_{L2}(G)=\frac{1}{M} \sum_{i=1}^M\left(\mathbf{y}_{i}-G(\mathbf{x}_i)\right)^2
\end{equation}
and the $\mathcal{L}(VF)$ is same as  \eqref{VF_representation}.

\subsection{Loss Functions of DDIM for Fluid Topology Optimization}
The objective of DDIM is to obtain the most realistic $x_0$, by training the parameter $\theta$ of the model. Negative log maximum likelihood estimation is used to compute the parameters of the diffusion model.  A simplified version of the loss function based on predicting noise $\varepsilon$ that is proposed in \cite{Dif-1-cD-ho2020denoising}    is used as the loss function for our framework. The loss function for DDIM is expressed as follows:

\begin{equation}
\mathcal{L_D}=E_{t, z_0, \varepsilon}\left[\| \varepsilon-\varepsilon_\theta\left(x_t, \mathbf{x}, t\right) \|^2\right]
\end{equation}
where $0 \leq t \leq T$ is the time step of adding noise from $x_0$, $\varepsilon_t$ is the added noise, and $\varepsilon_{\theta}$ is a neural network(function approximator of $\varepsilon$) used for predicting the adding noise from $x_0$ to $x_t$. In this problem, $\mathbf{x}$ is the additional channels as conditions input to the neural network $\varepsilon_{\theta}$.

\section{Numerical Experiments}
For the training process, the ADAM optimizer \cite{3-1kingma2014adam}, which is a gradient-based optimization algorithm for stochastic objective functions, is used to determine the optimal weights of the encoder and decoder. The neural network was trained on a computer with an Intel(R)Core(TM) i7-12700 CPU and an NVIDIA GeForce GTX1650 Ti GPU (RTX A5000 GPU for DDIM) using the Keras framework.\\
\subsection{Data Set Generation \label{Sec::DataSet}}

 For data set generation, as the first part of the study, we start by setting up the design domain and boundary conditions. Figure \ref{fig::RegionFst} shows the design domain we set up for experiments, where we consider the unit height of the design domain to be 1, and the unit width to be delta (we fix the width delta to be 1.5). The design domain is regularly discretized by 10000($100 \times 100$) rectangle elements, and a total of 10201 grid points are used for representing the physical fields and optimal results. The boundary condition distributions on the boundary of the design domain are shown in Figure \ref{fig::RegionFst}:
\begin{figure}
    \centering
    \includegraphics[width=9cm]{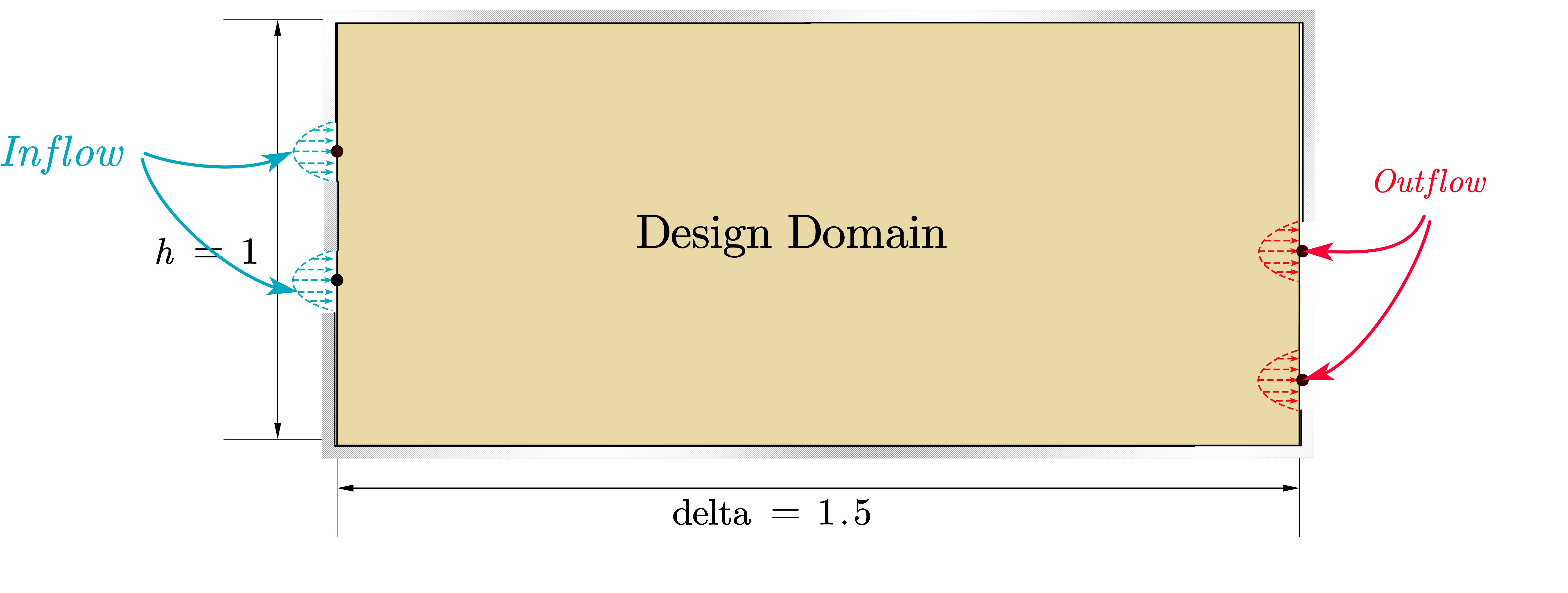}
    \caption{Design domain of samples}
    \label{fig::RegionFst}
\end{figure}

 First, the open-source topology optimization code \cite{5-borrvall2003topology, 5-1-guillaume2004topological, 5-2-aage2008topology},( Dolfin Adjoint framework for topology optimization of Stokes flow \cite{2-mitusch2019dolfin, 5-borrvall2003topology}), is used to create data sets of the topology optimization for fluid field together with optimization settings and boundary conditions. These data sets are separated into training sets and test sets. The training set is used for training the neural network model, and the test set is used for examining the performance of the NN model. Then, the neural network is trained based on the given training data set, and the mapping from input physical fields to its corresponding optimal result is constructed, as shown in Figure \ref{fig::CNN}.

In order to generate the mapping between input and output, we need to construct the datasets including the major information we need to encode into the input channel, and the optimal topology output for such given input information.

In order to examine the presented framework more comprehensively, two data sets are generated to evaluate the performance of the presented method. A total of $3500$ optimized fluid optimization results are generated for each data set by the source code \cite{5-borrvall2003topology, 5-1-guillaume2004topological, 5-2-aage2008topology}. We randomly select  $3000$ samples as the training set and $500$ as the test set.

\subsubsection{Data Set \uppercase\expandafter{\romannumeral1}}
\begin{figure}
    \centering
    \includegraphics[width=17cm]{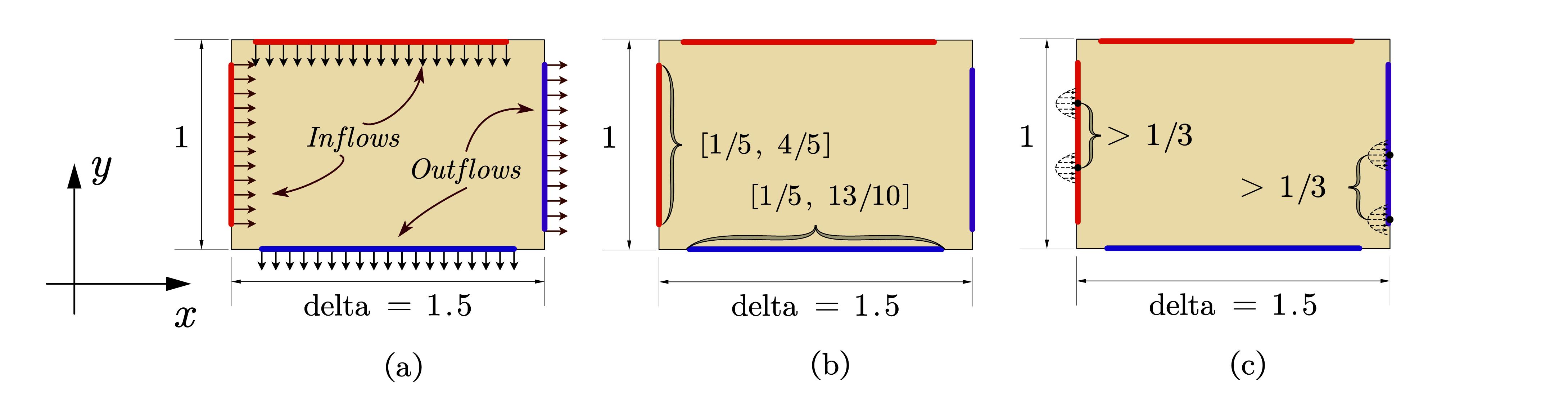}
    \caption{Boundary conditions for design domain of data set \uppercase\expandafter{\romannumeral1}}
    \label{fig::RegionBgin}
\end{figure}

The data set \uppercase\expandafter{\romannumeral1} is generated using the following conditions.
\begin{itemize}
\item The Inflows are distributed at the left and upper boundaries, as shown in Figure \ref{fig::RegionBgin} $(a)$
\item The Outflows are distributed at the right and lower boundaries, as shown in Figure \ref{fig::RegionBgin} $(a)$
\item The possible center of the Inflow lies on the left boundary y-axis $[1/5,4/5]$ and bottom boundary x-axis $[1/5,13/10]$, as shown in Figure \ref{fig::RegionBgin} $(b)$.
\item The possible center of the Outflow lies on the right boundary y-axis $[1/5,4/5]$ and upper boundary x-axis $[1/5,13/10]$, as shown in Figure \ref{fig::RegionBgin} $(b)$.
\item The width of the Inflow channel and that of the Outflow channel are the same, which is defined as $1/6$.
\item The number of Inflow channels are randomly allocated from 1 to 3 and the same number of Outflow channels are allocated.
\item The distance between the two center points of Inflows should be at least $1/3$, as shown in Figure \ref{fig::RegionBgin} $(c)$.
\item The distance between the two center points of Outflows should be at least $1/3$, as shown in Figure \ref{fig::RegionBgin} $(c)$.
\end{itemize}

Consider $\mathbf{x_{ij}}$ is the center point for each inflow channel, where $i$ stands for inflow, $j$ is the index of the inflow channel, which can be from 1 to 3. Similarly, $\mathbf{x_{oj}}$ is the center point of each outflow channel, where $o$ stands for the outflow. The velocity distribution for inflow and outflow boundary satisfies the polynomial, due to the property of Poiseuille flow. Thus, the velocity distribution of the inflow and outflow  are defined as follows:
 when the center position of the inflow is at the left boundary, the inflow center position is at the y-axis, and it can be represented as $\mathbf{x_{ij}} = (0, x_{ij})$. The corresponding velocity distribution is $\boldsymbol{u_{ij}}$, and it can be represented as
$\boldsymbol{u_{ij}} = (u_{i,j},0)$. The  velocity distribution is shown as equation \eqref{eq::infloweqn}. Similarly, when the center position is at the upper boundary, the inflow center position is represented as $\mathbf{x_{ij}} = (x_{ij}, 1)$, and the velocity distribution can be represented as $\boldsymbol{u_{ij}} = (0,u_{i,j})$. The  $u_{i,j}$ distribution is also shown as equation \eqref{eq::infloweqn}.
\begin{equation}\label{eq::infloweqn}
u_{i,j} =\bar{g} \cdot\left(1-\left(2 \frac{x-x_{i,j}}{l}\right)^2\right) \text { at }\left[x_{i,j}-\frac{l}{2}, x_{i,j}+\frac{l}{2}\right]
\end{equation}
For outflow, at the bottom boundary, the center position is  $\boldsymbol{x_{o,j}} = (x_{o,j}, 0)$
and the corresponding velocity: $\boldsymbol{u_{o,j}} = (0, u_{o,j})$; at right boundary, the center position is  $\boldsymbol{x_{o,j}} = (1.5, x_{o,j})$, the corresponding velocity $\boldsymbol{u_{o,j}} = (u_{o,j}, 0)$:

\begin{equation}
u_{o,k} =-\bar{g} \cdot\left(1-\left(2 \frac{x-x_{o,k}}{l}\right)^2\right) \text { at }\left[x_{o,k}-\frac{l}{2}, x_{o,k}+\frac{l}{2}\right]
\end{equation}
where $\bar{g}$ is corresponding to the maximum velocity, $l$ is the width of the flow channel.

\subsubsection{Data Set \uppercase\expandafter{\romannumeral2}}
\begin{figure}
    \centering
    \includegraphics[width=16cm]{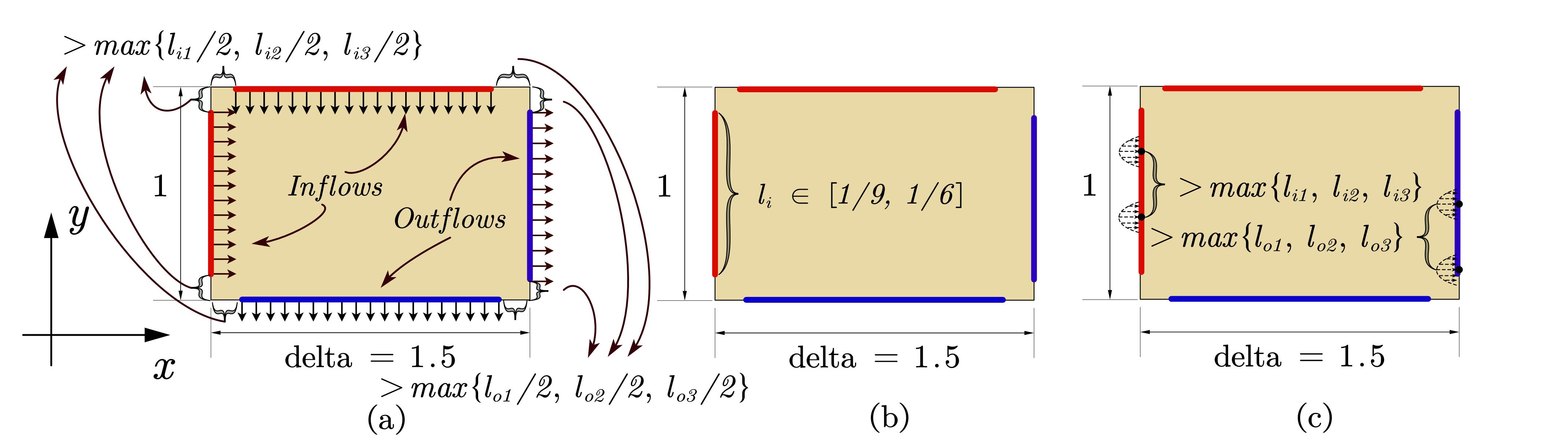}
    \caption{Boundary conditions for design domain of data set \uppercase\expandafter{\romannumeral2}}
    \label{fig::RegionBgin2}
\end{figure}

In order to examine the flexibility of our model, we generate data set \uppercase\expandafter{\romannumeral2}. Given Inflow channels with random width $l_{i,j}$,  and Outflow channels with random width $l_{o,k}$, the center position for each channel to be $x_{i,j}$ for Inflow, $x_{o.j}$ for Outflow, where $i$ and $o$ are the random numbers of Inflows and Outflows respectively, $j$ and $k$ are the flow index for Inflow and Outflow respectively, the boundary conditions of the design domain are generated as follows:
\begin{itemize}
\item The number of Inflow channels is randomly allocated from 1 to 3, and the number of outflow channels is randomly allocated from 1 to 3, which does not need to be the same as the number of inflow channels.
\item The Inflows are distributed at the left and upper boundaries, as shown in Figure \ref{fig::RegionBgin2} $(a)$,
\item The Outflows are distributed at the right and lower boundaries, as shown in Figure \ref{fig::RegionBgin2} $(a)$
\item The possible center of the Inflow lies on the left boundary y-axis $ \left[max\{l_{i,1}, l_{i,2}, l_{i,3}\}, 1 - max\{l_{i,1}, l_{i,2}, l_{i,3}\}\right]$ and x-axis $\left[max\{l_{i,1}, l_{i,2}, l_{i,3}\}, 1.5 - max\{l_{i,1}, l_{i,2}, l_{i,3}\}\right]$ bottom boundary, as shown in Figure \ref{fig::RegionBgin2} $(a)$.
\item The possible center of the Outflow lies on the right boundary y-axis $\left[max\{l_{o,1}, l_{o,2}, l_{o,3}\}, 1 -  max\{l_{o,1}, l_{o,2}, l_{o,3}\}\right]$ and x-axis $[max\{l_{o,1}, l_{o,2}, l_{o,3}\}, 1.5 - max\{l_{o,1}, l_{o,2}, l_{o,3}\}]$ upper boundary, as shown in Figure \ref{fig::RegionBgin2} $(a)$.
\item The width of the Inflow channels and Outflow channels are randomly set from $1/9$ to $1/6$. The difference between the widest and narrowest channels is set not to exceed $20\%$, as shown in Figure \ref{fig::RegionBgin2} $(b)$
\item The distance between the center points of two channels is set to be no less than the width of the largest channel as shown in Figure \ref{fig::RegionBgin2} $(c)$.
\end{itemize}

The velocity distribution $\boldsymbol{u_{i,j}}$ for Inflow at the boundary can be represented as $\boldsymbol{u_{i,j}} = u_{i,j} \boldsymbol{\hat{n}}$, where $\hat{n}$ is the normal vector of the boundary:
\begin{equation}
u_{i,j} = k_1  \cdot\left(1-\left(2 \frac{x-x_{i,j}}{l}\right)^2\right) \text { at }\left[x_{i,j}-\frac{l_{i,j}}{2}, x_{i,j}+\frac{l_{i,j}}{2}\right]
\end{equation}

\begin{equation}
u_{o,k} = - k_2  \cdot\left(1-\left(2 \frac{x-x_{o,k}}{l}\right)^2\right) \text { at }\left[x_{o,k}-\frac{l_{o,k}}{2}, x_{o,k}+\frac{l_{o,k}}{2}\right]
\end{equation}
where $k_1$ and $k_2$ are determined by the random total flux passing through the domain which will be further computed in \eqref{Eq::k1k2}.

The volume fraction(VF) is set up to be positively correlated with the total width of all channels of the Inflow and Outflow, and the number of channels for Inflow and Outflow.

\begin{equation}
    VF = \left( 12 - \frac{|O| + |I|}{12} \right) \cdot \sum_{j = 1}^{|I|} l_{i,j} + \sum_{k = 1}^{|O|} l_{o,k}
\end{equation}
where $I$ is the Inflow width set, $O$ is the Outflow width set, i.e. $I = \{l_{i,1}, l_{i,2}, l_{i,3}\}$,  $O = \{l_{o,k}, l_{o,2}, l_{o,3}\}$.

Considering the total flux passing through the design domain is Q,  to make the model more generalizable, we randomly select the total flux Q to be:
\begin{equation}
    Q = \left(|O| + |I|\right) \cdot 2 \cdot random\{1/9, 1/6 \}
\end{equation}
where $random\{1/9, 1/6 \}$ denotes the value is uniformly selected from $1/9$ to $1/6$.  If we denote the 
$Q_{Inflow}$ to be the total Inflow flux and $Q_{Outflow}$ be the total Outflow flux, combined with their speed distribution at the input and output ports, they can each be expressed as:  
\begin{equation}\label{eq::Q}
\begin{aligned}
& Q_{\text {Inflow}}=\sum_{j=i}^{|I|} \int_{-\frac{l_{i, j}}{2}}^{\frac{l_{i, j}}{2}} k_1\left(-x^2+\frac{l_{i, j}^2}{4}\right) d x \\
& Q_{\text {Outflow}}=\sum_{k=1}^{\mid O|} \int_{-\frac{l_{o, k}}{2}}^{\frac{l_{o, k}}{2}} k_2\left(-x^2+\frac{l_{o, k}}{4}\right) d x
\end{aligned}
\end{equation}
Since the Stokes flow in the design domain is incompressible, total flux passing into the design domain is the same as the flux passing out of the domain, that is:
\begin{equation}
    Q = Q_{\text{Input}} = Q_{\text{Output}}
\end{equation}
In \eqref{eq::Q}, $x, l, Q$ are known, therefore $k_1$ and $k_2$ can be obtained by:
\begin{equation} \label{Eq::k1k2}
\begin{aligned}
& k_1=\frac{6 Q_{\text{Outflow}}}{\sum_{j = 1}^{|I|} x_{i,j}} \\
& k_2=\frac{6 Q_{\text{Input}}}{\sum_{k = 1}^{|O|} x_{o,k}}
\end{aligned}
\end{equation}
Hence, the Inflow and Outflow are determined.
\subsection{Evaluation Metrics}
To evaluate the performance of the presented method, we use two commonly used metrics: mean square error(MSE) and mean absolute error(MAE). Denote the predicted result to be  $\hat{\mathbf{y}}=NN(f(x))$, the optimization results calculated using typical methods are used as ground truth which is denoted as $\mathbf{y}$. The performance of the presented cGAN for fluid topology optimization is evaluated by comparing $\hat{\mathbf{y}}$ and $\mathbf{y}$. Both prediction $\hat{\mathbf{y}}$ and the ground truth $\mathbf{y}$ are of the shape $M \times M$, where $M$ is $101$. The average absolute differences between the prediction values and the ground truth are measured by MAE. Considering there are N samples in a data set,  the MAE for such data set is:
\begin{equation}\label{eq::MAE}
\text { MAE }=\frac{1}{N} \sum_{i=1}^N\left|\mathbf{y}_{i}-\hat{\mathbf{y}}_{i}\right|
\end{equation}
MSE measures the average squared difference between the estimated values and the ground truth. For a data set with $N$ samples, the MSE is given by:
\begin{equation}\label{eq::MSE}
\mathrm{MSE}=\frac{1}{M} \sum_{i=1}^M\left(\mathbf{y}_{i}-\hat{\mathbf{y}}_{i}\right)^2
\end{equation}
\subsection{Comparison and Selection of Physical Fields}

The basic flow features and the extended features including spatial gradients are mentioned in Table \ref{table::designvariables}. Different input channels might result in different results for neural networks. Thus, various physical field combinations mentioned in Table \ref{table::designvariables} are examined in this study as input for our presented neural network. In order to select the best input combination of channels, we compare the performance of the neural network by comparing MAE and MSE of each group of combinations of input channels under the same training sample and test sample. The comparison of MAE and MSE for each channel combination is shown in Table \ref{Tab::Combi_input}. In Table \ref{Tab::Combi_input}, the first column is the sample number, the second column is the physical field used by the sample, and the third to sixth columns are the MAE and MSE of the training and test sets, respectively.

\begin{table*}
\centering
\caption{Comparison optimal designs between physical fielData Set \label{Tab::Combi_input}}
\begin{tabular}{|c|c|c|c|c|c|}
\hline \multicolumn{2}{|c|}{ Metrics } & \multicolumn{2}{|c|}{ MAE } & \multicolumn{2}{|c|}{ MSE } \\
\hline № & Physical Fields & Training  & Test & Training & Test \\
\hline 0 & $\mathrm{VF}+v+u+p$  & 0.008011 & 0.036026 & 0.002829 & 0.008837 \\
\hline 1 & $\mathrm{VF}+|U|$ & 0.008023 & 0.040075 & 0.002989 & 0.012389 \\
\hline 2 & $\mathrm{VF}+u + v + p + |U| + \frac{\partial u}{\partial x} + \frac{\partial u}{\partial x}+ \frac{\partial u}{\partial y}+ \frac{\partial v}{\partial x}+ \frac{\partial v}{\partial y} + \frac{\partial p}{\partial x} + \frac{\partial p}{\partial y}$ & 0.008646 & 0.033405 & 0.003143 & 0.008257\\
\hline 3 & $\mathrm{VF}+ \frac{\partial p}{\partial x} + \frac{\partial p}{\partial y}$ & 0.010625 & 0.041667 & 0.004432 & 0.012994 \\
\hline 4 & $\mathrm{VF}+v+u$  & 0.010443 & 0.034866 & 0.004266 & 0.008983 \\
\hline 5 & $\mathrm{VF}+ |U| + p$ & 0.010352 & 0.038937 & 0.004218 & 0.011584 \\
\hline 6 & $\mathrm{VF}+\frac{\partial u}{\partial x} + \frac{\partial u}{\partial x}+ \frac{\partial u}{\partial y}+ \frac{\partial v}{\partial x}+ \frac{\partial v}{\partial y}$ & 0.008719 & 0.035922  & 0.003143 & 0.008719\\
\hline 7 & $\mathrm{VF}+|U| + p +  \frac{\partial p}{\partial x} + \frac{\partial p}{\partial y}$ & 0.007620 & 0.037946 & 0.002737 & 0.011117 \\
\hline 8 & $\mathrm{VF}+ |U| +u + v + \frac{\partial u}{\partial x} + \frac{\partial u}{\partial x}+ \frac{\partial u}{\partial y}+ \frac{\partial v}{\partial x}+ \frac{\partial v}{\partial y}$ & 0.008649 & 0.034601& 0.003154 & 0.008263 \\
\hline 9 & $\mathrm{VF}+ p +u + v  + \frac{\partial u}{\partial x} + \frac{\partial u}{\partial x}+ \frac{\partial u}{\partial y}+ \frac{\partial v}{\partial x}+ \frac{\partial v}{\partial y}$ & 0.008595 & 0.033748& 0.002858 & 0.008542 \\
\hline
\end{tabular}
\end{table*}
The results shown in Table \ref{Tab::Combi_input} indicate that the combination № $2$ has the best performance on the test set. Thus, we try to use  $\mathrm{VF}+u + v + p + |U| + \frac{\partial u}{\partial x} + \frac{\partial u}{\partial x}+ \frac{\partial u}{\partial y}+ \frac{\partial v}{\partial x}+ \frac{\partial v}{\partial y} + \frac{\partial p}{\partial x} + \frac{\partial p}{\partial y}$ as our input channel.

\subsection{Performance and Model Evaluation}
The presented data-driven architecture for solving fluid topology optimization is examined on the three neural networks introduced in Section \ref{Sec::Methods}: CNN, cGAN, and DDIM respectively. The performance of each neural network framework is evaluated by pixel-wise errors. For each neural network, the comparison between the generated results and the corresponding density-based results is shown to evaluate performance subjectively.
\subsubsection{CNN based UNet Framework}
\begin{figure}
\centering
\begin{minipage}{.45\textwidth}
  \centering
  \includegraphics[width=0.9\linewidth]{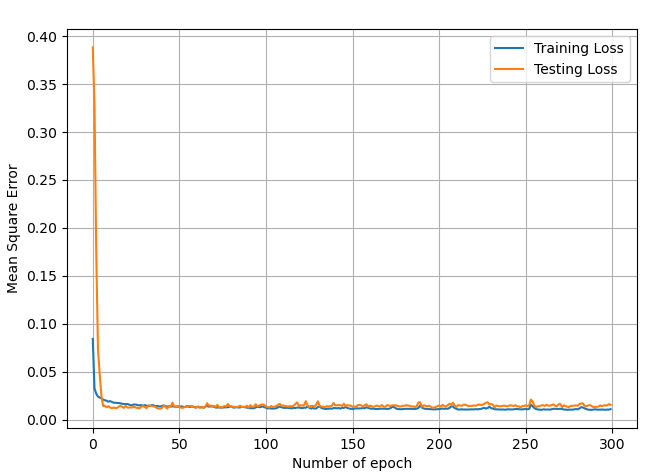}
  \captionof{figure}{MAE of CNN in training}
  \label{fig:CNNtest1Loss}
\end{minipage}
\begin{minipage}{.45\textwidth}
  \centering
  \includegraphics[width=0.9\linewidth]{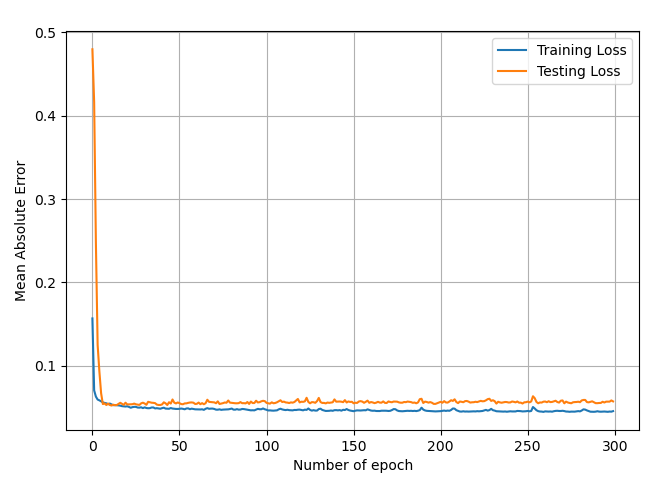}
  \captionof{figure}{MSE of CNN in training}
  \label{fig:CNNtest2Loss}
\end{minipage}
\end{figure}

The first framework that we examined is the UNet-structured CNN, as shown in Figure \ref{fig::CNN}. 
We use the data set \uppercase\expandafter{\romannumeral1} and data set \uppercase\expandafter{\romannumeral2} generated in \ref{Sec::DataSet} to evaluate the performance of the fluid topology optimization framework using CNN. Using the $\mathrm{VF}+u + v + p + |U| + \frac{\partial u}{\partial x} + \frac{\partial u}{\partial x}+ \frac{\partial u}{\partial y}+ \frac{\partial v}{\partial x}+ \frac{\partial v}{\partial y} + \frac{\partial p}{\partial x} + \frac{\partial p}{\partial y}$ as the input channel, 
the training loss for mean absolute error (MAE) and mean squared error (MSE) via epochs for the training set and the test set are presented, in Figure \ref{fig:CNNtest1Loss} and \ref{fig:CNNtest2Loss} respectively. 

\begin{figure}
    \centering
    \includegraphics[width=13.4cm]{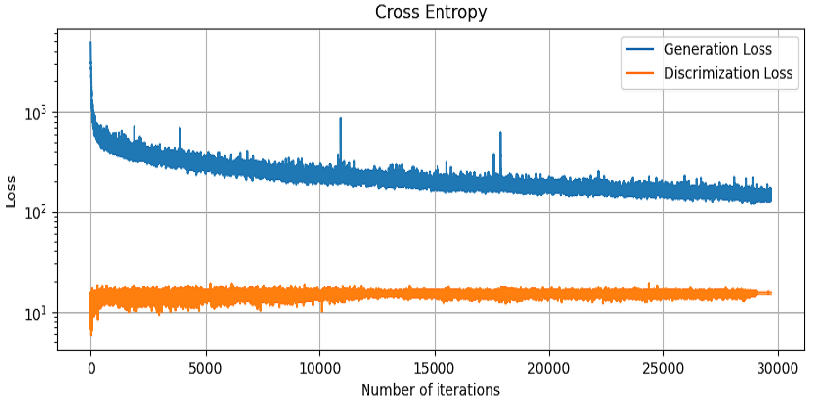}
    \caption{Loss of discriminator of cGAN framework, and generator of cGAN framework during training}
    \label{fig::iterationAndLoss}
\end{figure}

\begin{figure}
\centering
\subfloat[NN result for Data Set \uppercase\expandafter{\romannumeral1}]{\label{sfig:a}\includegraphics[width=3.0cm,height=2.1cm]{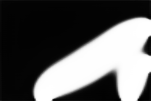}}\hfill
\subfloat[GT result for Data Set \uppercase\expandafter{\romannumeral1}]{\label{sfig:b}\includegraphics[width=3.0cm,height=2.1cm]{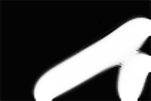}}\hfill
\subfloat[NN result for Data Set \uppercase\expandafter{\romannumeral2}]{\label{sfig:c}\includegraphics[width=3.0cm,height=2.1cm]{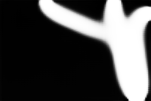}}\hfill
\subfloat[GT result for Data Set \uppercase\expandafter{\romannumeral2}]{\label{sfig:e}\includegraphics[width=3.0cm,height=2.1cm]{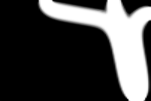}}\\
\subfloat[NN result 2 in Data Set \uppercase\expandafter{\romannumeral1}]
{\label{sfig:a}\includegraphics[width=3.0cm,height=2.1cm]{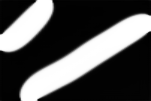}}\hfill
\subfloat[GT result 2 in Data Set \uppercase\expandafter{\romannumeral1}]{\label{sfig:b}\includegraphics[width=3.0cm,height=2.1cm]{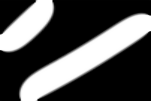}}\hfill
\subfloat[NN result 2 in Data Set \uppercase\expandafter{\romannumeral2}]{\label{sfig:c}\includegraphics[width=3.0cm,height=2.1cm]{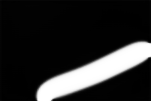}}\hfill
\subfloat[GT result 2 in Data Set \uppercase\expandafter{\romannumeral2}]{\label{sfig:e}\includegraphics[width=3.0cm,height=2.1cm]{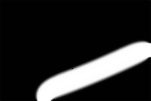}}\\
\subfloat[NN result 3 in Data Set \uppercase\expandafter{\romannumeral1}]
{\label{sfig:a}\includegraphics[width=3.0cm,height=2.1cm]{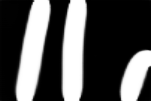}}\hfill
\subfloat[GT result 3 in Data Set \uppercase\expandafter{\romannumeral1}]{\label{sfig:b}\includegraphics[width=3.0cm,height=2.1cm]{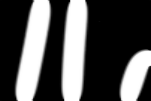}}\hfill
\subfloat[NN result 3 in Data Set \uppercase\expandafter{\romannumeral2}]{\label{sfig:c}\includegraphics[width=3.0cm,height=2.1cm]{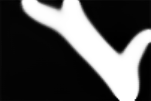}}\hfill
\subfloat[GT result 3 in Data Set \uppercase\expandafter{\romannumeral2}]{\label{sfig:e}\includegraphics[width=3.0cm,height=2.1cm]{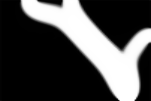}}\\
\subfloat[NN result 4 in Data Set \uppercase\expandafter{\romannumeral1}]
{\label{sfig:a}\includegraphics[width=3.0cm,height=2.1cm]{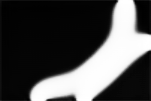}}\hfill
\subfloat[GT result 4 in Data Set \uppercase\expandafter{\romannumeral1}]{\label{sfig:b}\includegraphics[width=3.0cm,height=2.1cm]{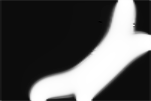}}\hfill
\subfloat[NN result 4 in Data Set \uppercase\expandafter{\romannumeral2}]{\label{sfig:c}\includegraphics[width=3.0cm,height=2.1cm]{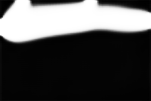}}\hfill
\subfloat[GT result 4 in Data Set \uppercase\expandafter{\romannumeral2}]{\label{sfig:e}\includegraphics[width=3.0cm,height=2.1cm]{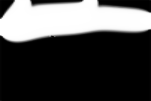}}\\
\subfloat[NN result 5 in Data Set \uppercase\expandafter{\romannumeral1}]
{\label{sfig:a}\includegraphics[width=3.0cm,height=2.1cm]{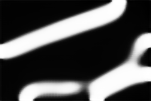}}\hfill
\subfloat[GT result 5 in Data Set \uppercase\expandafter{\romannumeral1}]{\label{sfig:b}\includegraphics[width=3.0cm,height=2.1cm]{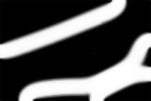}}\hfill
\subfloat[NN result 5 in Data Set \uppercase\expandafter{\romannumeral2}]{\label{sfig:c}\includegraphics[width=3.0cm,height=2.1cm]{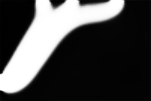}}\hfill
\subfloat[GT result 5 in Data Set \uppercase\expandafter{\romannumeral2}]{\label{sfig:e}\includegraphics[width=3.0cm,height=2.1cm]{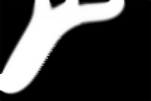}}\\
\captionsetup{width=1.0 \linewidth}
\caption{The 5 generated structures through CNN framework randomly selected from the test set of Data Set \uppercase\expandafter{\romannumeral1}  and \uppercase\expandafter{\romannumeral2}, and the corresponding optimal structure computed through density-based method}
\label{fig:CNNRestest}
\end{figure}
The performance of CNN is illustrated in Figure \ref{fig:CNNRestest}. The $5$ generated structures through CNN framework are randomly selected from the test set of Data Set I and II. The first(third) column in Figure \ref{fig:CNNRestest} shows the generated result from CNN for Data Set I(II); the second(fourth) column of Figure \ref{fig:CNNRestest} shows the result computed through the density-based method for Data Set \uppercase\expandafter{\romannumeral1}(\uppercase\expandafter{\romannumeral2}).
\subsubsection{cGAN Framework}
Model Evaluation for Data Set \uppercase\expandafter{\romannumeral1}:
The cross-entropy loss of cGAN for solving the Stokes problem is shown in Figure \ref{fig::iterationAndLoss}, where the blue line is the generator loss while training, and the orange line is the discriminator loss while training. It can be seen that the generator loss first declines quickly, and then goes stable, while the discriminator loss first increases, and then goes stable.

\begin{figure}
\subfloat[MSE of the cGAN framework in training for Data Set \uppercase\expandafter{\romannumeral1}]{\label{fig:test2cGAN}\includegraphics[width=8.4cm,height=5.775cm]{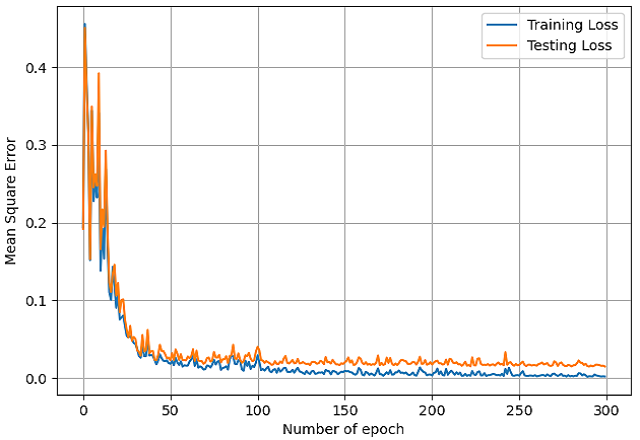}}\hfill
\subfloat[MAE of the cGAN framework in training for Data Set \uppercase\expandafter{\romannumeral1}]{\label{fig:test1cGAN}\includegraphics[width=8.4cm,height=5.775cm]{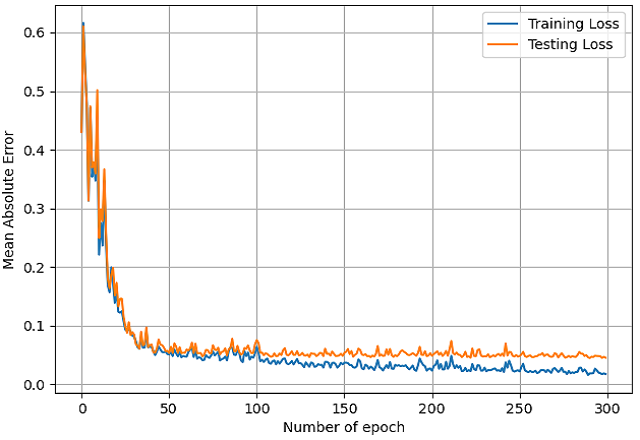}}\\
\subfloat[MSE of the cGAN framework in training for Data Set \uppercase\expandafter{\romannumeral2}]
{\label{fig:uneqtest2}\includegraphics[width=8.4cm,height=5.775cm]{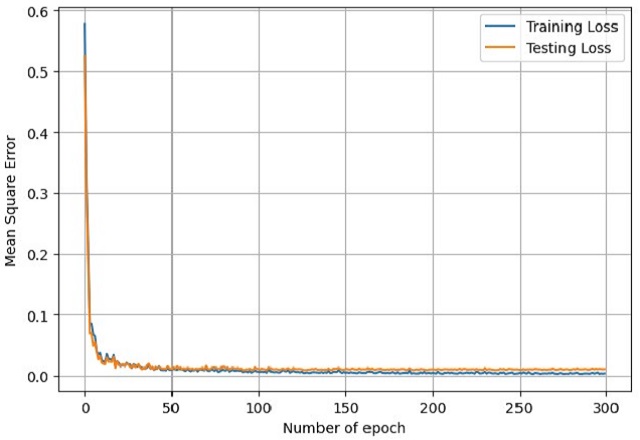}}\hfill
\subfloat[MAE of the cGAN framework in training for Data Set \uppercase\expandafter{\romannumeral2}]{\label{fig:uneqtest1}\includegraphics[width=8.4cm,height=5.775cm]{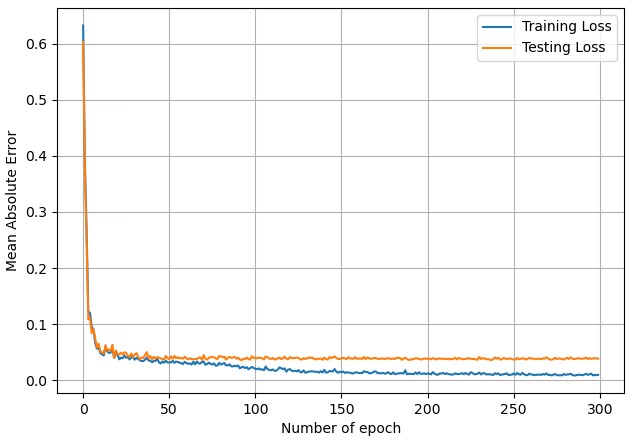}}\hfill
\caption{MAE and MSE of cGAN in training}
\end{figure}

The history of loss on the training and test data sets for the data set \uppercase\expandafter{\romannumeral1} during the training procedure are shown in Figure \ref{fig:test1cGAN} and Figure \ref{fig:test2cGAN}. The validation loss for mean absolute error is stable, with convergence after 200 epochs while the training loss still goes down and the over-fitting appears. However, the test loss does not further increase, while the training loss decreases while over-fitting. Therefore, the over-fitting is not a serious problem.

Model evaluation of cGAN for Data Set \uppercase\expandafter{\romannumeral2}:
The MAE loss and MSE loss for data set \uppercase\expandafter{\romannumeral2} using cGAN to generate the result for fluid topology optimization are shown in Figure \ref{fig:uneqtest1} and Figure \ref{fig:uneqtest2} respectively. while training is more stable compared with that of the first data set as shown in Figure \ref{fig:test1cGAN} and Figure \ref{fig:test2cGAN}. This neural network structure is also effective in calculating optimization results for data set \uppercase\expandafter{\romannumeral2}. Since
data set \uppercase\expandafter{\romannumeral2} is more flexible than data set \uppercase\expandafter{\romannumeral1}, our presented neural network has some generalization capability.

\begin{figure}
\subfloat[NN result for Data Set \uppercase\expandafter{\romannumeral1}]{\label{sfig:a}\includegraphics[width=3.0cm,height=2.1cm]{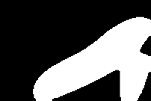}}\hfill
\subfloat[GT result for Data Set \uppercase\expandafter{\romannumeral1}]{\label{sfig:b}\includegraphics[width=3.0cm,height=2.1cm]{GroundTruthDataSetOne0.png}}\hfill
\subfloat[NN result for Data Set \uppercase\expandafter{\romannumeral2}]{\label{sfig:c}\includegraphics[width=3.0cm,height=2.1cm]{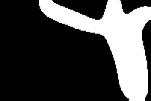}}\hfill
\subfloat[GT result for Data Set \uppercase\expandafter{\romannumeral2}]{\label{sfig:e}\includegraphics[width=3.0cm,height=2.1cm]{GroundTruthDataSetTwo0.png}}\\
\subfloat[NN result 2 in Data Set \uppercase\expandafter{\romannumeral1}]
{\label{sfig:a}\includegraphics[width=3.0cm,height=2.1cm]{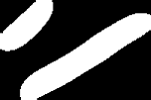}}\hfill
\subfloat[GT result 2 in Data Set \uppercase\expandafter{\romannumeral1}]{\label{sfig:b}\includegraphics[width=3.0cm,height=2.1cm]{GroundTruthDataSetOne1.png}}\hfill
\subfloat[NN result 2 in Data Set \uppercase\expandafter{\romannumeral2}]{\label{sfig:c}\includegraphics[width=3.0cm,height=2.1cm]{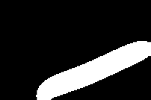}}\hfill
\subfloat[GT result 2 in Data Set \uppercase\expandafter{\romannumeral2}]{\label{sfig:e}\includegraphics[width=3.0cm,height=2.1cm]{GroundTruthDataSetTwo1.png}}\\
\subfloat[NN result 3 in Data Set \uppercase\expandafter{\romannumeral1}]
{\label{sfig:a}\includegraphics[width=3.0cm,height=2.1cm]{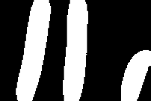}}\hfill
\subfloat[GT result 3 in Data Set \uppercase\expandafter{\romannumeral1}]{\label{sfig:b}\includegraphics[width=3.0cm,height=2.1cm]{GroundTruthDataSetOne2.png}}\hfill
\subfloat[NN result 3 in Data Set \uppercase\expandafter{\romannumeral2}]{\label{sfig:c}\includegraphics[width=3.0cm,height=2.1cm]{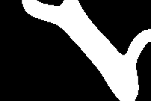}}\hfill
\subfloat[GT result 3 in Data Set \uppercase\expandafter{\romannumeral2}]{\label{sfig:e}\includegraphics[width=3.0cm,height=2.1cm]{GroundTruthDataSetTwo2.png}}\\
\subfloat[NN result 4 in Data Set \uppercase\expandafter{\romannumeral1}]
{\label{sfig:a}\includegraphics[width=3.0cm,height=2.1cm]{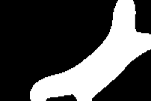}}\hfill
\subfloat[GT result 4 in Data Set \uppercase\expandafter{\romannumeral1}]{\label{sfig:b}\includegraphics[width=3.0cm,height=2.1cm]{GroundTruthDataSetOne3.png}}\hfill
\subfloat[NN result 4 in Data Set \uppercase\expandafter{\romannumeral2}]{\label{sfig:c}\includegraphics[width=3.0cm,height=2.1cm]{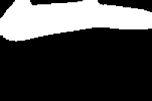}}\hfill
\subfloat[GT result 4 in Data Set \uppercase\expandafter{\romannumeral2}]{\label{sfig:e}\includegraphics[width=3.0cm,height=2.1cm]{GroundTruthDataSetTwo3.png}}\\
\subfloat[NN result 5 in Data Set \uppercase\expandafter{\romannumeral1}]
{\label{sfig:a}\includegraphics[width=3.0cm,height=2.1cm]{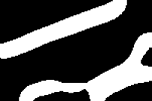}}\hfill
\subfloat[GT result 5 in Data Set \uppercase\expandafter{\romannumeral1}]{\label{sfig:b}\includegraphics[width=3.0cm,height=2.1cm]{GroundTruthDataSetOne4.png}}\hfill
\subfloat[NN result 5 in Data Set \uppercase\expandafter{\romannumeral2}]{\label{sfig:c}\includegraphics[width=3.0cm,height=2.1cm]{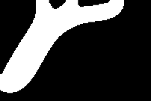}}\hfill
\subfloat[GT result 5 in Data Set \uppercase\expandafter{\romannumeral2}]{\label{sfig:e}\includegraphics[width=3.0cm,height=2.1cm]{GroundTruthDataSetTwo4.png}}\\
\captionsetup{width=1.0\linewidth}
\caption{The 5 generated structures through cGAN framework randomly selected from the test set of Data Set \uppercase\expandafter{\romannumeral1}   \uppercase\expandafter{\romannumeral2}, and the corresponding optimal structure computed through density-based method}
\label{fig::cGANResultsTestSet}
\end{figure}

Figure \ref{fig::cGANResultsTestSet} shows the visual comparison between the results obtained by the density-based method and the results generated by cGAN. The performance of the presented cGAN for data set \uppercase\expandafter{\romannumeral1}  is illustrated in the left two columns of Figure \ref{fig::cGANResultsTestSet}.  
 By comparing each sample, we find that for most samples, the presented cGAN method can generate accurate results. While, for some structures (i.e. Figure \ref{fig::cGANResultsTestSet} (q)), inaccuracies and errors may occur locally.

The performance of the presented cGAN neural network for Data Set \uppercase\expandafter{\romannumeral2} is illustrated in the right two columns of Figure \ref{fig::cGANResultsTestSet}.  On the test set of Data Set \uppercase\expandafter{\romannumeral2}, cGAN gives accurate results on most cases except the first one(Figure \ref{fig::cGANResultsTestSet} (c)).

We subjectively find that the cGAN framework performs better than that of the CNN framework on both Data Set \uppercase\expandafter{\romannumeral1} and Data Set \uppercase\expandafter{\romannumeral2} by comparing results for CNN in Figure \ref{fig:CNNRestest}, and results for cGAN in Figure \ref{fig::cGANResultsTestSet}.
\subsubsection{Diffusion Model Framework}
\begin{figure}
\subfloat[NN result for Data Set \uppercase\expandafter{\romannumeral1}]{\label{sfig:a}\includegraphics[width=3.0cm,height=2.1cm]{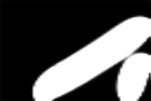}}\hfill
\subfloat[GT result for Data Set \uppercase\expandafter{\romannumeral1}]{\label{sfig:b}\includegraphics[width=3.0cm,height=2.1cm]{GroundTruthDataSetOne0.png}}\hfill
\subfloat[NN result for Data Set \uppercase\expandafter{\romannumeral2}]{\label{sfig:c}\includegraphics[width=3.0cm,height=2.1cm]{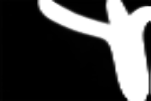}}\hfill
\subfloat[GT result for Data Set \uppercase\expandafter{\romannumeral2}]{\label{sfig:e}\includegraphics[width=3.0cm,height=2.1cm]{GroundTruthDataSetTwo0.png}}\\
\subfloat[NN result 2 in Data Set \uppercase\expandafter{\romannumeral1}]
{\label{sfig:a}\includegraphics[width=3.0cm,height=2.1cm]{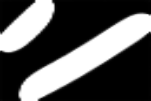}}\hfill
\subfloat[GT result 2 in Data Set \uppercase\expandafter{\romannumeral1}]{\label{sfig:b}\includegraphics[width=3.0cm,height=2.1cm]{GroundTruthDataSetOne1.png}}\hfill
\subfloat[NN result 2 in Data Set \uppercase\expandafter{\romannumeral2}]{\label{sfig:c}\includegraphics[width=3.0cm,height=2.1cm]{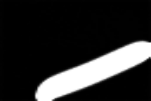}}\hfill
\subfloat[GT result 2 in Data Set \uppercase\expandafter{\romannumeral2}]{\label{sfig:e}\includegraphics[width=3.0cm,height=2.1cm]{GroundTruthDataSetTwo1.png}}\\
\subfloat[NN result 3 in Data Set \uppercase\expandafter{\romannumeral1}]
{\label{sfig:a}\includegraphics[width=3.0cm,height=2.1cm]{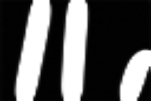}}\hfill
\subfloat[GT result 3 in Data Set \uppercase\expandafter{\romannumeral1}]{\label{sfig:b}\includegraphics[width=3.0cm,height=2.1cm]{GroundTruthDataSetOne2.png}}\hfill
\subfloat[NN result 3 in Data Set \uppercase\expandafter{\romannumeral2}]{\label{sfig:c}\includegraphics[width=3.0cm,height=2.1cm]{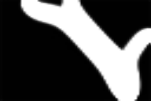}}\hfill
\subfloat[GT result 3 in Data Set \uppercase\expandafter{\romannumeral2}]{\label{sfig:e}\includegraphics[width=3.0cm,height=2.1cm]{GroundTruthDataSetTwo2.png}}\\
\subfloat[NN result 4 in Data Set \uppercase\expandafter{\romannumeral1}]
{\label{sfig:a}\includegraphics[width=3.0cm,height=2.1cm]{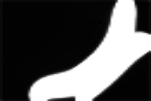}}\hfill
\subfloat[GT result 4 in Data Set \uppercase\expandafter{\romannumeral1}]{\label{sfig:b}\includegraphics[width=3.0cm,height=2.1cm]{GroundTruthDataSetOne3.png}}\hfill
\subfloat[NN result 4 in Data Set \uppercase\expandafter{\romannumeral2}]{\label{sfig:c}\includegraphics[width=3.0cm,height=2.1cm]{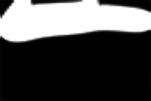}}\hfill
\subfloat[GT result 4 in Data Set \uppercase\expandafter{\romannumeral2}]{\label{sfig:e}\includegraphics[width=3.0cm,height=2.1cm]{GroundTruthDataSetTwo3.png}}\\
\subfloat[NN result 5 in Data Set \uppercase\expandafter{\romannumeral1}]
{\label{sfig:a}\includegraphics[width=3.0cm,height=2.1cm]{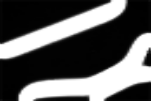}}\hfill
\subfloat[GT result 5 in Data Set \uppercase\expandafter{\romannumeral1}]{\label{sfig:b}\includegraphics[width=3.0cm,height=2.1cm]{GroundTruthDataSetOne4.png}}\hfill
\subfloat[NN result 5 in Data Set \uppercase\expandafter{\romannumeral2}]{\label{sfig:c}\includegraphics[width=3.0cm,height=2.1cm]{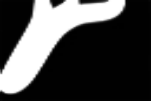}}\hfill
\subfloat[GT result 5 in Data Set \uppercase\expandafter{\romannumeral2}]{\label{sfig:e}\includegraphics[width=3.0cm,height=2.1cm]{GroundTruthDataSetTwo4.png}}\\
\captionsetup{width=1.0\linewidth}
\caption{The 5 generated structures through DDIM framework randomly selected from the test set of Data Set \uppercase\expandafter{\romannumeral1}   \uppercase\expandafter{\romannumeral2}, and the corresponding optimal structure computed through density-based method}
\label{fig::DiffusionResTest1}
\end{figure}

To evaluate the subjective performance of using diffusion models on generating the result of topology optimization for the fluid channel, we selected 5 random samples from the test set of Data Set \uppercase\expandafter{\romannumeral1}, and Data Set \uppercase\expandafter{\romannumeral2} respectively (same random samples as that of cGAN and CNN frameworks). Figure \ref{fig::DiffusionResTest1} shows the comparison of ground truth which is the optimal result obtained with the SIMP method and the result generated from the diffusion model framework. There are small errors in some structures, and overall, the subjective performance is similar to that of cGAN but better than CNN. 



\begin{figure}
    \centering
        \begin{minipage}{.5\textwidth}
          \centering
          \includegraphics[width=2.7in]{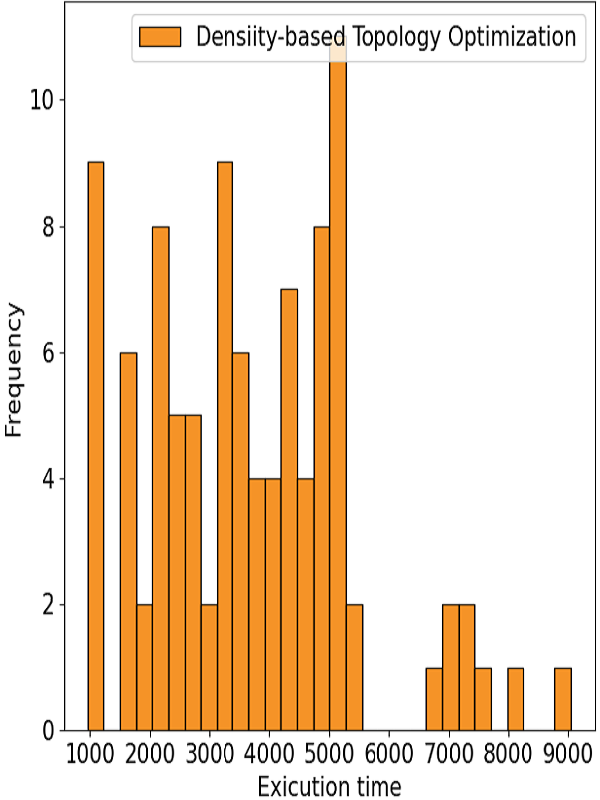}
          \captionsetup{width=0.9\linewidth}
          \captionof{figure}{Execution time of traditional framework while sampling}
          \label{fig:TimeSIMP}
        \end{minipage}%
        \begin{minipage}{.5\textwidth}
          \centering
          \includegraphics[width=2.7in]{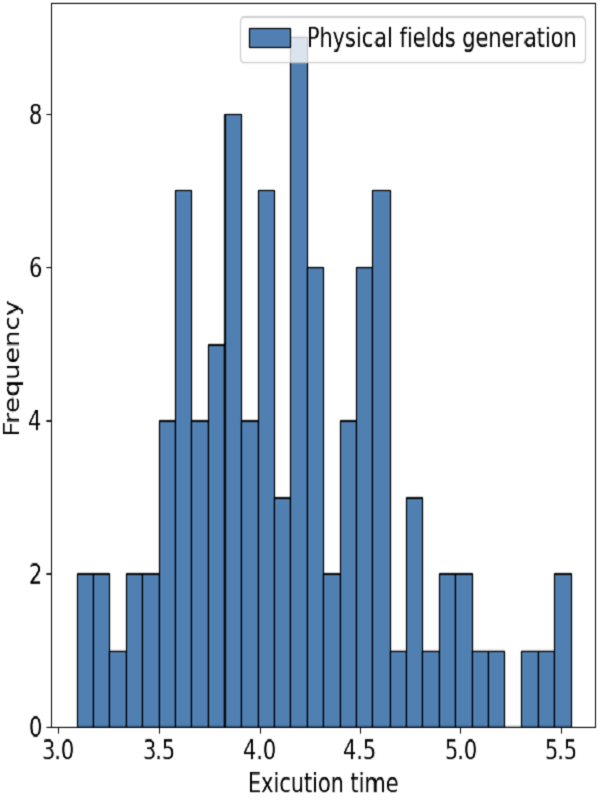}
          \captionsetup{width=0.9\linewidth}
          \captionof{figure}{Execution time of physical fields generating while sampling}
          \label{fig:TimePhysicalFieldsGeneration}
        \end{minipage}
\end{figure}
\subsubsection{Accuracy Comparison}
The objective pixel-wise loss comparison for the presented CNN framework, cGAN framework, and DDIM framework are summarized in Table \ref{Table::ComparecGANandCNN}. For both MAE error and MSE error, the loss of cGAN on the test set and training set is lower than that of the CNN and DDIM framework. 
\begin{table}
\caption{The final loss comparison between cGAN, CNN and DDIM\label{Table::ComparecGANandCNN}}
\centering
\begin{tabular}{|c|c|c|c|c|}
\hline \multirow{2}{*}{ Model } & \multicolumn{2}{|c|}{ MAE } & \multicolumn{2}{c|}{ MSE } \\
\cline { 2 - 5 } & Training & Testing & Training & Testing \\
\hline CNN & 0.038673 & 0.055308 & 0.010624 & 0.021393 \\
cGAN & \textbf{0.008646} & \textbf{0.033405} & \textbf{0.003143} & \textbf{0.008257} \\
DDIM & 0.027836 & 0.037274 & 0.003796 & 0.009128 \\
\hline
\end{tabular}
\end{table}
\subsection{Exicution Time Comparison}
\begin{figure}
\subfloat[Execution time for CNN]{\label{TimeNN:a}\includegraphics[width=5.4cm,height=8.0cm]{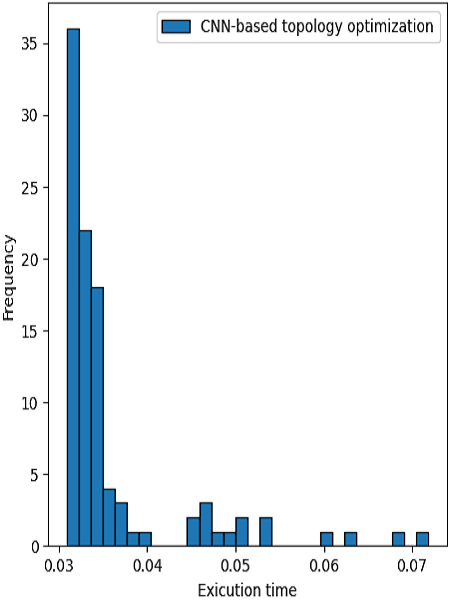}}\hfill
\subfloat[Execution time of cGAN]{\label{TimeNN:b}\includegraphics[width=5.4cm,height=8.0cm]{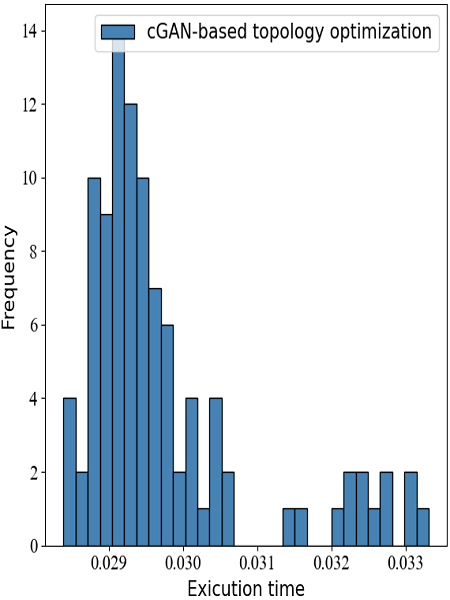}}\hfill
\subfloat[Execution time for DDIM]{\label{TimeNN:c}\includegraphics[width=5.4cm,height=8.0cm]{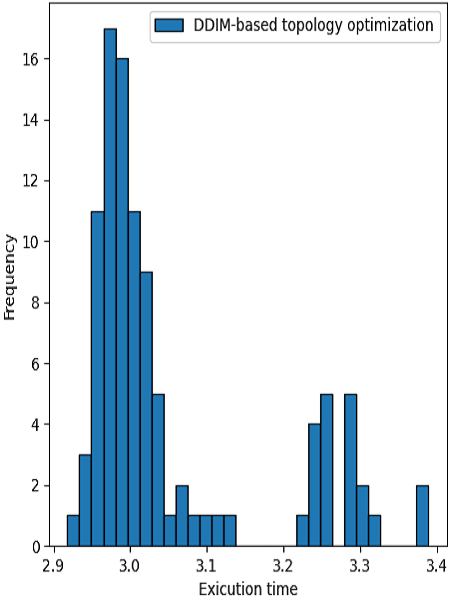}}\\
\captionsetup{width=1.0\linewidth}
\caption{Execution time for topology optimization using neural network}
\label{fig:TimeNN}
\end{figure}

In this section, we compare the execution time of the density-based method with that of the presented data-driven method for fluid topology optimization.
We randomly generated 100 sets of initial boundary conditions in  Data Set \uppercase\expandafter{\romannumeral2}, all of which satisfy our dataset boundary condition settings in Section \ref{Sec::DataSet}. We first performed the topology optimization using the traditional density-based approach. The maximum number of iterative loops was chosen to be 50, and the time taken to compute each sample was counted. The execution time distribution for the density-based method is shown in Figure \ref{fig:TimeSIMP}, where the horizontal axis is the execution time of the topology optimization using the traditional density-based method, and the vertical axis is the number of samples during the certain interval of execution time. For different samples, due to the randomly assigned boundary conditions, the convergence speed of the traditional density-based method may be different due to the different working conditions, so the time required to obtain the optimal structure for different samples varies greatly. The execution time of each sample is mainly in the range from 1600 to 3000 seconds. The average execution time for computing each sample by the density-based method is 2653.46 seconds. On the other hand, solving fluid topology optimization using the presented data-driven method is far faster than the density-based method.
Solving fluid topology optimization through the presented data-driven method is divided into two parts:
The first part is to compute the physical fields on the original, unoptimized design domain and the second part is the computed optimal structure through neural network mapping from physical fields to the optimal structure. 
In the first part,  the execution time distribution of the generated physical fields is shown in Figure \ref{fig:TimePhysicalFieldsGeneration}. From Figure \ref{fig:TimePhysicalFieldsGeneration}, it can be seen that the execution is relatively stable, where the execution time is mainly distributed around 4 seconds, with an average computation time of 3.6588 seconds. In the second part, the optimal structure is calculated by the neural network, and the distribution of execution time, which is referred to as the execution time of the neural network, is shown in Figure \ref{TimeNN:a}, Figure \ref{TimeNN:b}, and Figure \ref{TimeNN:c} for CNN, cGAN and DDIM models respectively.
For CNN and cGAN framework, the execution time for computing optimal results through mapping is also stable, with an average of about 0.037 seconds. The execution time for sampling using CNN and cGAN is negligible compared with the execution time of part 1. For DDIM framework, the average execution time for sampling is 3.053 seconds. 
 
In summary, the execution time of data-driven fluid topology optimization is about 3.6958 seconds on average for CNN and cGAN frameworks, and 6.711 seconds on average for the DDIM framework, which is negligible compared with the execution time of the density-based method for solving fluid topology optimization, which is 2653.46 seconds on average. 

\section{Conclusion and Future Work}
We develop the data-driven method to solve the topology optimization of the channel flow problems. The presented method is based on three conditional-generated frameworks including UNet, cGAN, and DDIM, and the generated results are examined on two data sets respectively. The data-driven method for topology optimization of the channel flow problems achieves a huge time reduction in the average execution time compared with the traditional density-based method. Meanwhile, there is a tiny difference in accuracy between the results obtained through data-driven topology optimization and those obtained through the density-based method.\\
~\\
The limitation of the presented method is that the data-driven method takes a long time to generate a large enough data set to satisfy the best performance of the neural network, and the trained neural network model is limited to working on a specific design domain. When the shape of the design domain modifies, the samples need to be regenerated through the new design domain, and the network needs to be retrained based on the new sample. This issue might be solved by generating a comprehensive dataset while adding more input channels to the neural network to enhance the generalization of the neural network. In the future work, the design domain can be encoded and added as a new channel to the input information of the neural network. In addition, we only study the results for the Stokes flows. More general fluidic problems, such as Navier-Stokes flow topology optimizations and other kinds of fluid-based topology optimizations need to be further investigated.

\section*{Acknowledgements}
This research was funded by CAS Project for Young Scientists in Basic Research (YSBR-$066$), the National Natural Science Foundation of China (Nos. $51875545$), the Innovation Grant of Changchun Institute of Optics, Fine Mechanics and Physics (CIOMP), and the Science and Technology Development Plan of Jilin Province (No. $2023008783$).

\bibliographystyle{unsrt}  
\bibliography{references}

\end{document}